\documentclass[10pt]{article}
\usepackage{amsmath,amscd}
\usepackage{amsfonts}
\usepackage{amssymb}
\usepackage{pictexwd,dcpic}

\begin{document}
\author{Yves Barmaz \thanks{Section de Math\'ematiques,
Universit\'e de Gen\`eve,
2-4 rue du Li\`evre,
c.p. 64, 1211 Gen\`eve 4,
Switzerland, yves.barmaz@unige.ch} }
\title{T-duality through BV Morphisms and BV Pushforwards in Topological Field Theories}
\maketitle
\begin{abstract}
We introduce the concept of duality between quantum field theories in the Batalin-Vilkovisky formalism, which is interpreted either as a BV morphism, the result of dual BV pushforwards or a combination of both. When a BV morphism affects only the target space of a given model, we call it T-duality. To justify this name, we demonstrate how topological aspects of T-duality in string theory such as the relation between curvature and H-flux or isomorphisms of Courant algebroids are equivalent to dualities of topological sigma models in two and three dimensions.
\end{abstract}
\section{Introduction}
T-duality (short for target space duality) was first discovered within the framework of string theory, where two theories compactified on circles (or more generally on tori) are equivalent under certain conditions. If the geometry of the target space is given by a principal circle bundle endowed with background metric and H-flux, gauging of the $S^1$ symmetry and integration of the newly established gauge connection leads to a T-dual model on a different circle bundle with different background fields, related to the ones of the initial theory through the Buscher rules \cite{Buscherrules1}, \cite{Buscherrules2}. The geometrical and topological content was formalized by Bouwknegt, Mathai and others \cite{Bouwknegt1}, \cite{Bouwknegt2}, \cite{Bouwknegt3}. The main result relates the H-flux of a model with the first Chern class of the circle bundle geometry of its T-dual. From there on, one can show that the twisted cohomologies on the target spaces of the two T-dual theories are isomorphic. Furthermore, the $S^1$ invariant differential forms underlying these cohomologies can be interpreted as Clifford modules for certain Courant algebroids, so it is possible to find an isomorphism of the Courant algebroids that is compatible with this isomorphism of the Clifford modules \cite{TdualityCourant}. In the realm of generalized geometry, T-duality is therefore defined as an isomorphism of two Courant algebroids in a purely mathematical way, that is without reference to any physical model, and several results can be re-derived independently on the initial physical theory. For instance, the Buscher rules can be deduced from the behavior of a generalized metric on the Courant algebroid under this T-duality isomorphism.

In parallel, these topological structures have found themselves at the heart of several topological field theories (models that do not depend on a metric), such as the two-dimensional twisted Poisson sigma model \cite{PSMfirst} \cite{WZPSM} or the three-dimensional Courant sigma model \cite{AKSZCSM}. It might thus be tempting to repeat the procedure initially used to derive the Buscher rules on these models and retrieve only the topological content of T-duality. However, the price to pay for the topological nature of these models is a complicated set of symmetries (the kinetic terms of string sigma models with background fields break these symmetries) that renders their action degenerate. The Batalin-Vilkovisky formalism is thus ideally suited to deal with these theories. In this formalism, the space of classical fields is extended with so-called ghosts and antifields to a much richer BV space of fields equipped with an odd symplectic form (the BV structure) and a BV Laplacian. Quantization then requires the classical action to be extended to a BV action that has to satisfy a master equation. Even though the BV machinery is usually hard to apply, it provides us with a natural method to construct effective theories (through so-called ``BV pushforwards'') and a nice interpretation of T-duality as either a BV morphism, namely a symplectomorphism with respect to the BV structures, or the result of dual BV pushforwards. Our goal is therefore to express the topological aspects of T-duality as BV dualities of \emph{ad hoc} topological models: the twisted Poisson sigma model with a trivial Poisson structure, which describes the topological sector of the sigma model for a string theory with background fields, and the Courant sigma model, built from a Courant algebroid following the AKSZ prescription.

We begin our exposition with a short introduction to the basic aspects of the BV formalism and the AKSZ construction in section \ref{BVformalism}. In section \ref{BVdualities}, we describe three different approaches to dualities offered by the BV formalism. As a first example, we show in section \ref{TdualityCSM} how a BV morphism applied to a certain type of Courant sigma models reproduces an isomorphism of Courant algebroids first constructed by Cavalcanti and Gualtieri from geometric considerations. In section \ref{Tduality}, we briefly review the main results regarding T-duality in physics and geometry in order to understand the motivation behind this isomorphism and to provide material for our second example involving dual BV pushforwards, that we treat in section \ref{TdualityPSM}. Finally, we extend this discussion to the case of general principal torus bundles in section \ref{TorusBundles}.
\paragraph{Acknowledgment}
The author would like to thank Anton Alekseev, Alberto Cattaneo and Pavol \v{S}evera for remarks and comments, Pavel Mnev for his valuable insights into the BV formalism, Maxim Zabzine for a useful discussion at an early stage of this project and Peter Bouwknegt for introducing him to T-duality.

This research was supported by the SNF grant number 140985.
\section{The Batalin-Vilkovisky Formalism} \label{BVformalism}
\subsection{Basics}
The Batalin-Vilkovisky formalism \cite{BVtheorem} was developed in the 80's as a tool for the quantization of degenerate quantum field theories, for instance gauge theories. Given a theory described in the Lagrangian formalism by a classical action $S_{\mathrm{cl}}$ defined over a space $\mathcal{F}_{\mathrm{cl}}$ of classical fields,  the mathematical data for its classical BV formulation, the so-called \emph{classical BV manifold}, contains three elements $(\mathcal{F},\Omega, S)$. First, a BV space of fields $\mathcal{F}$, which is a $\mathbb{Z}$-graded (or sometimes $\mathbb{Z}_2$-graded) infinite-dimensional manifold, extends the space $\mathcal{F}_{\mathrm{cl}}$ of classical fields. The internal degree associated to this grading is called ``ghost number'' and is set to zero for the classical fields. Second, this BV space of fields carries a symplectic structure $\Omega$ of ghost number $-1$ (the BV structure), whose induced odd Poisson bracket $\left\lbrace\cdot,\cdot\right\rbrace$ has ghost number $1$ (the BV bracket). Third, a BV action $S$ defined on $\mathcal{F}$ extends the classical action in the sense that $S\vert_{\mathcal{F}_{\mathrm{cl}}} = S_{\mathrm{cl}}$ and satisfies the classical master equation
\begin{displaymath}
\left\lbrace S, S\right\rbrace =0.
\end{displaymath}
As a consequence of this classical master equation, the Hamiltonian vector field generated by $S$, $Q=\left\lbrace S, \cdot \right\rbrace$, is cohomological, $\left[ Q, Q \right]=0$.

If $\mathcal{F}$ is furthermore equipped with an integration measure $\mu$, we can define a Laplacian operator $\Delta$ (of degree $1$) acting on a function $f$ on $\mathcal{F}$ as
\begin{displaymath}
\Delta f = \mathrm{div}_\mu \left\lbrace f, \cdot \right\rbrace,
\end{displaymath}
namely the divergence with respect to $\mu$ of the Hamiltonian vector field generated by $f$. To define the quantum BV theory, we need a measure $\mu$ such that this Laplacian squares to zero, $\Delta^2 =0$. We will call the measures with this property \emph{BV measures}. If this condition is satisfied, a theorem by Batalin and Vilkovisky \cite{BVtheorem} states that the integral over a Lagrangian submanifold $\mathcal{L}$ of $\mathcal{F}$ of a given function $f$ on $\mathcal{F}$ is constant under continuous deformations of $\mathcal{L}$,
\begin{displaymath}
\int_{\mathcal{L}} \sqrt{\mu}_{\mathcal{L}} \, f = \int_{\mathcal{L'}} \sqrt{\mu}_{\mathcal{L'}} \, f,
\end{displaymath}
provided $f$ is BV-harmonic, $\Delta f =0$. Here $\sqrt{\mu}_{\mathcal{L}}$ is the measure on $\mathcal{L}$ induced by $\mu$, and $\mathcal{L}'$ is a deformation of $\mathcal{L}$. Furthermore, the integral of the BV Laplacian of a function vanishes,
\begin{displaymath}
\int_{\mathcal{L}} \sqrt{\mu}_{\mathcal{L}} \, \Delta g =0.
\end{displaymath}

The problem with a degenerate theory is that integrals such as the one of the partition function
\begin{displaymath}
Z = \int_{\mathcal{F}_{\mathrm{cl}}} \sqrt{\mu}_{\mathrm{cl}} \, e^{\frac{i}{\hbar}S_{\mathrm{cl}}}
\end{displaymath}
do not make sense due to the degeneracy. As a remedy, we can find a Lagrangian submanifold $\mathcal{L}_{\mathrm{cl}}$ of $\mathcal{F}$ that contains $\mathcal{F}_{\mathrm{cl}}$, a BV measure $\mu$ on $\mathcal{F}$ that reproduces $\mu_{\mathrm{cl}}$ when restricted to $\mathcal{F}_{\mathrm{cl}}$, and finally a quantum BV action $W$ on $\mathcal{F}$ such that $W\vert_{\mathcal{L}_{\mathrm{cl}}}=S_{\mathrm{cl}}$ and
\begin{equation} \label{QME1}
\Delta e^{\frac{i}{\hbar}W} = 0,
\end{equation}
along with a Lagrangian submanifold $\mathcal{L}$ deformed from $\mathcal{L}_{\mathrm{cl}}$ on which $W$ is non-degenerate, and the Batalin-Vilkovisky theorem ensures that
\begin{equation}
Z = \int_{\mathcal{F}_{\mathrm{cl}}} \mu_{\mathrm{cl}} \, e^{\frac{i}{\hbar}S_{\mathrm{cl}}} = \int_{\mathcal{L}}  \sqrt{\mu}_{\mathcal{L}} \, e^{\frac{i}{\hbar}W},
\end{equation}
where the expression on the right-hand side can be computed, usually perturbatively.

The difficult task is to find a solution $W$ of the quantum master equation (\ref{QME1}), which is equivalent to
\begin{equation}
\frac{1}{2}\left\lbrace W,W\right\rbrace = i\hbar \, \Delta W,
\end{equation}
where the relation with the classical master equation is obvious. In practice, one first looks for a solution $S$ of the classical master equation, and then proceeds to solve the QME order by order in $\hbar$, with the expansion $W=S+\sum_{k=1}^\infty \hbar^k W_k$, provided no anomaly prevents the existence of a solution. In other words, we work with formal power series in $\hbar$ of functions, $W \in \mathrm{Fun}(\mathcal{F})\left[\left[ \hbar \right]\right]$.

The mathematical data for the quantum BV formulation of a QFT is the \emph{quantum BV manifold} $(\mathcal{F},\Omega,\mu, W)$, which contains one more element than its classical counterpart, namely a BV measure $\mu$, and where the classical BV action $S$ satisfying the classical master equation has been replaced by the quantum BV action $W$ satisfying the quantum master equation (see for instance \cite{SchwarzGeomQME} for details).

Note that the BV Laplacian is singular, and its proper definition requires the use of some regularization procedure. Nevertheless, in the theories we are dealing with in this paper, the BV Laplacian can be regularized in such a way that the solution $S$ of the CME will be BV harmonic and will thus already satisfy the QME.

Finally, the algebra of observables in the quantum BV formalism $\mathcal{O}_{\mathrm{quant}}$ is defined as the subset of functions $f$ on $\mathcal{F}$ that satisfy the condition
\begin{equation}
\Delta (f e^{\frac{i}{\hbar}W}) = 0,
\end{equation}
which is equivalent to
\begin{equation}
i\hbar\Delta f + \left\lbrace f, W \right\rbrace = 0.
\end{equation}
This allows us to define their expectation value in a similar way as the partition function,
\begin{equation}
\langle f \rangle = \frac{1}{Z} \int_{\mathcal{L}}  \sqrt{\mu}_{\mathcal{L}} \, f \, e^{\frac{i}{\hbar}W}.
\end{equation}
Notice that in the classical limit, we obtain the condition
\begin{equation}
\left\lbrace f, S \right\rbrace = 0
\end{equation}
for the algebra of classical observables $\mathcal{O}_{\mathrm{clas}}$, and that quantum observables may also be constructed as formal power series in $\hbar$, starting at zeroth order with a classical observable. 
\subsection{The AKSZ Construction} \label{AKSZconstr}
\subsubsection{Generalities}
While it is usually hard to find the BV extension of a given classical degenerate action, the AKSZ construction \cite{AKSZconstruction}, based on geometrical considerations, leads to solutions of the classical master equation that sometimes involve interesting models, such as the Chern-Simons theory or the Poisson sigma model used to derive Kontsevich's formula for deformation quantization \cite{AKSZPSM}. This construction is well suited to implement T-duality, as we will see that a symplectomorphism of the target space of an AKSZ model can be naturally lifted to a full BV morphism of the AKSZ space of fields. The AKSZ construction has been extensively treated in the literature, so here we will just give a brief explanation of the Courant sigma model \cite{AKSZCSM}, that we will use to illustrate a T-duality BV morphism in the AKSZ formalism.

We know that due to the classical master equation, the Hamiltonian (with respect to the BV structure) vector field $Q=\left\lbrace S,\cdot\right\rbrace$ generated by the BV action $S$ is cohomological. The idea behind the AKSZ construction is therefore to build a cohomological vector field on a graded symplectic manifold, and see if it is Hamiltonian.

The AKSZ space of fields consists of maps from $T\left[1\right]\Sigma_{n+1}$, the tangent bundle of some $(n+1)$-dimensional closed\footnote{The case of manifolds with boundaries requires a careful treatment of the boundary conditions or can be treated in the BV-BFV formalism of Cattaneo, Mnev and Reshetikhin \cite{BVBFV}.} manifold $\Sigma_{n+1}$ with the degree of its fibers shifted by one, to a graded symplectic manifold $Y$ equipped with a symplectic structure $\omega_Y$ of degree $n$ (this is the internal degree, also called ghost number, as opposed to the degree as a differential form, which is of course $2$) and a cohomological Hamiltonian vector field $Q_Y$ generated by a function $\Theta_Y$ of degree $n+1$ on $Y$, i.e. $\imath_{Q_Y} \omega_Y = \delta \Theta_Y$, where $\delta$ denotes the exterior derivative on $Y$. Moreover, we assume that a Liouville form $\alpha_Y$ is associated to $\omega_Y=\delta\alpha_Y$. Differential forms on the target space (such as $\omega_Y$, $\alpha_Y$ or $\Theta_Y$) can be lifted to the AKSZ space of fields
\begin{displaymath}
\mathcal{F} = \mathrm{Map}(T\left[1\right]\Sigma_{n+1}, Y)
\end{displaymath}
via a pullback by the evaluation map
\begin{displaymath}
\mathrm{ev}: T\left[1\right]\Sigma_{n+1} \times \mathcal{F} \rightarrow Y
\end{displaymath}
followed by an integration on the source space $T\left[1\right]\Sigma_{n+1}$ (see \cite{AKSZPSM} for details). For the integration of a function on an odd tangent bundle, one normally uses the canonical measure $\mu$. These functions are identified with differential forms on the base manifold $\Sigma_{n+1}$, and the top-form is extracted and integrated over $\Sigma_{n+1}$.

Explicitly, the AKSZ BV structure is defined as
\begin{equation}
\Omega = \int_{T\left[1\right]\Sigma_{n+1}} \mu \, \mathrm{ev}^\ast( \omega_Y).
\end{equation}
Note that the exterior derivative on $Y$ becomes the exterior variational derivative in the space of fields, which explains the choice of $\delta$ for its notation.

Before we can give the AKSZ action, we need to define the de Rham vector field on $\mathcal{F}$,
\begin{displaymath}
Q_D = \sum_a D\phi^a \frac{\delta}{\delta\phi^a},
\end{displaymath}
where the sum runs over all the (super)fields of $\mathcal{F}$ (the fields $\phi^a$ can be interpreted as coordinate-fields of $\mathcal{F}$) and $D$ is the de Rham vector field on the source space. Explicitly, if $u^\mu, \ \mu=1,\dots,n+1$ are some coordinates on $\Sigma_{n+1}$ and $\theta^\mu$ the corresponding odd coordinates (with ghost number 1) along the fibers of $T\left[1\right]\Sigma_{n+1}$, we define the de Rham vector field on $T\left[1\right]\Sigma_{n+1}$ as $D=\theta^\mu \frac{\partial}{\partial u^\mu}$.

In the AKSZ action, the kinetic term is constructed by contracting $Q_D$ with the pullback by the evaluation map of the Liouville potential $\alpha_Y$, and the interaction term simply with $\Theta_Y$,
\begin{equation}
S = \int_{T\left[1\right]\Sigma_{n+1}} \mu \left( \imath_{Q_D} \mathrm{ev}^\ast(\alpha_Y) + \mathrm{ev}^\ast(\Theta_Y) \right).
\end{equation}

We mention in passing that if $n=1$, we can set $Y=T^\ast\left[1\right]M$ for some manifold $M$ with $\omega$ being the canonical symplectic structure, and we obtain the AKSZ construction of the Poisson sigma model.
\subsubsection{The Courant Sigma Model} \label{CSM}
Of interest in this paper is the case $n=2$, which leads to the so-called Courant sigma model, based on a Courant algebroid.

We recall that a Courant algebroid over a manifold $M$ is a vector bundle $\mathcal{E}$ over $M$ equipped with a fiber-wise non-degenerate symmetric scalar product $\langle\cdot,\cdot\rangle_{\mathcal{E}}:\mathcal{E}\times\mathcal{E}\rightarrow M\times\mathbb{R}$, a bracket of sections $\left[\cdot,\cdot\right]:\Gamma\mathcal{E}\times\Gamma\mathcal{E}\rightarrow\Gamma\mathcal{E}$ and an anchor map $\rho:\mathcal{E}\rightarrow TM$ satisfying the axioms
\begin{equation} \label{Courantalgebroidaxioms}
\begin{array}{rcl}
\left[ \phi,\left[\chi,\psi\right]\right] &=& \left[\left[ \phi,\chi\right],\psi\right] + \left[ \chi,\left[\phi,\psi\right]\right], \\
\left[\phi,f\psi\right] &=& \rho(\phi)f\,\psi + f\left[\phi,\psi\right], \\
\left[\phi,\phi\right] &=& \frac{1}{2}D\langle\phi,\phi\rangle, \\
\rho(\phi)\langle\psi,\psi\rangle &=& 2\langle\left[\phi,\psi\right],\psi\rangle,
\end{array}
\end{equation}
where $\phi,\psi,\chi$ are sections of $\mathcal{E}$ and $f$ a smooth function on $M$, and $D=\kappa\circ\rho^T\circ d$ with $\kappa:\mathcal{E}^\ast\rightarrow\mathcal{E}$ being the isomorphism induced by the inner product.

Given a closed three-form $H$, the direct sum $TM\oplus T^\ast M$ of a tangent and cotangent bundles of the same manifold $M$ equipped with the canonical product
\begin{displaymath}
\langle X + \alpha, Y + \beta\rangle = \frac{1}{2}(\alpha(Y)+\beta(X))
\end{displaymath}
and the Courant bracket
\begin{displaymath}
\left[X + \alpha, Y + \beta\right]_H = \left[X,Y\right] + \mathcal{L}_X \beta - \mathcal{L}_Y \alpha - \frac{1}{2}d(\imath_X \beta - \imath_Y \alpha) + \imath_X \imath_Y H
\end{displaymath}
provides the standard example of a Courant algebroid, called an exact Courant algebroid, since the sequence
\begin{displaymath}
0 \rightarrow T^\ast M \xrightarrow{\rho^\ast} \mathcal{E} \xrightarrow{\rho} TM \rightarrow 0
\end{displaymath}
is in this case exact, where we introduced the dual map of $\rho$, $\rho^\ast: T^\ast M \rightarrow \mathcal{E}^\ast \simeq \mathcal{E}$, and used the invariant scalar product on $\mathcal{E}$ to identify it with its dual. Furthermore, it has been shown  \cite{PavolClass} that any Courant algebroid that fits in such an exact sequence is isomorphic to $TM\oplus T^\ast M$ twisted by some closed three-form $H$, and that the corresponding equivalence classes are in one-to-one correspondence with the third de Rham cohomology classes of these twists, $\left[H\right] \in H^3(M)$, called \v{S}evera's classes.

Back to the AKSZ construction, if $\mathcal{E}$ is a Courant algebroid over $M$, we need to take as the target space $Y$ a subbundle of $T^\ast\left[2\right]\mathcal{E}\left[1\right]$ that corresponds to the isometric embedding $\mathcal{E}\hookrightarrow\mathcal{E}^\ast\oplus\mathcal{E}$ with respect to the Courant algebroid inner product on the left-hand side and the canonical product on the right-hand side. If $\mathcal{E}$ is an exact Courant algebroid, we can simply take $Y=T^\ast\left[2\right]T^\ast\left[1\right]M$.

In the superfield formalism, the field content of this model is given by a base map $X\in\mathrm{Map}(T\left[1\right]N\rightarrow M)$ and sections $p$ of the pullback of the shifted cotangent bundle $X^\ast T^\ast\left[2\right]M$ and $\Xi$ of the pullback of the shifted Courant algebroid $X^\ast \mathcal{E}\left[1\right]$. The space of fields supports the canonical BV structure
\begin{equation} \label{AKSZbv}
\Omega = \int_{T\left[1\right]N} \mu\, \left( \langle \delta p,\delta X\rangle + \frac{1}{2}\langle \delta\Xi,\delta\Xi\rangle_{\mathcal{E}}\right),
\end{equation}
where the first product is the canonical pairing between tangent and cotangent vectors. The AKSZ construction leads to the action
\begin{equation} \label{AKSZaction}
S = \int_{T\left[1\right]N} \mu\, \left( \langle p,DX\rangle + \frac{1}{2}\langle \Xi,D\Xi\rangle_{\mathcal{E}} - \langle p,\rho(\Xi)\rangle + \langle \Xi,\left[\Xi,\Xi\right]_{\mathcal{E}}\rangle_{\mathcal{E}} \right),
\end{equation}
built with the constructing blocks of a Courant algebroids: the anchor map, the Courant bracket and the fiber scalar product. One can show that the classical master equation is satisfied if and only if these elements satisfy the integrability conditions (\ref{Courantalgebroidaxioms}), and that Courant algebroids are uniquely encoded (up to Courant algebroid isomorphisms) in these Courant sigma models \cite{AKSZCSM}.

We illustrate the computations with the somewhat simpler (but relevant) case of an exact Courant algebroid $T^\ast M\oplus TM$ twisted by $H$. First, we may decompose the section $\Xi$ of $X^\ast \mathcal{E}\left[1\right]$ into its tangent and cotangent parts, $\xi$ and $\Theta$ respectively. If $M$ locally admits coordinates $x^i$, we can use them to express the superfields, $X^i$, $\Theta_i$, $\xi^i$ and $p_i$ of ghost number 0, 1, 1 and 2 respectively.  It is straightforward to write the BV structure
\begin{equation}
\Omega = \int_{T\left[1\right]N} \mu\, \left( \delta p_i\,\delta X^i - \delta\xi^i\,\delta\Theta_i \right)
\end{equation}
and the action
\begin{equation}
S = \int_{T\left[1\right]N} \mu\, \left( p_i(DX^i - \xi^i) + \frac{1}{2}\xi^i\,D\Theta_i + \frac{1}{2}\Theta_i\,D\xi^i + \frac{1}{6}H_{ijk}(X)\xi^i\xi^j\xi^k\right).
\end{equation}
Since integration along the odd dimensions of $T\left[1\right]N$ lowers the ghost number by $\mathrm{dim}(N)=3$, the BV structure has a ghost number $-1$ and the BV action  $0$ as expected.

To verify the classical master equation, we first compute the functional derivatives of the BV action in the superfield formalism. Due to different commutation rules, we need to define left- and right-derivatives separately. The trick is to compute the exterior derivative in the space of fields $\mathcal{F}$. If we denote by $\phi^a$ a generic superfield in $\mathcal{F}$ and assume that $a$ runs over all of them, we may define these derivatives as
\begin{equation} \label{functionalderivativessuperfields}
\delta S = \int_{T\left[1\right]N} \mu\, \sum_a \delta \phi^a \frac{\overrightarrow{\delta} S}{\delta \phi^a} = \int_{T\left[1\right]N} \mu\, \sum_a \frac{S \overleftarrow{\delta}}{\delta \phi^a} \delta \phi^a.
\end{equation}
Note that in the superfield formalism with an odd-dimensional $N$ on the source side and a functional of even ghost number (such as an action), $\delta \phi^a$ always commutes with the corresponding functional derivative, and we have
\begin{displaymath}
\frac{\overrightarrow{\delta} S}{\delta \phi^a} =  \frac{S \overleftarrow{\delta}}{\delta \phi^a},
\end{displaymath}
so we could as well drop the small arrows.

Explicitly, we find
\begin{displaymath}
\begin{array}{rcl}
\frac{\overrightarrow{\delta}S}{\delta p_i} &=& DX^i - \xi^i, \\
\frac{\overrightarrow{\delta}S}{\delta X^i} &=& - Dp_l + \frac{1}{6}\partial_l H_{ijk}(X)\xi^i\xi^j\xi^k, \\
\frac{\overrightarrow{\delta}S}{\delta \Theta_i} &=& D\xi^i, \\
\frac{\overrightarrow{\delta}S}{\delta \xi^i} &=& D\Theta_i - p_i + \frac{1}{2} H_{ijk}(X)\xi^j\xi^k,
\end{array}
\end{displaymath}
where we had to perform a few integrations by parts, using the fact that $\partial N=\emptyset$.

Finally, to compute the BV bracket of $S$ with itself, we need to invert the BV structure, which we can do in the superfield formalism,
\begin{displaymath}
\begin{split}
\frac{1}{2}\left\lbrace S,S \right\rbrace &= \int_{T\left[1\right]N} \mu\, \left( \frac{S\overleftarrow{\delta}}{\delta X^i}\frac{\overrightarrow{\delta}S}{\delta p_i} - \frac{S\overleftarrow{\delta}}{\delta \Theta_i}\frac{\overrightarrow{\delta}S}{\delta \xi^i} \right) \\
&= \int_{T\left[1\right]N} \mu\, \left( \left( - Dp_l + \frac{1}{6}\partial_l H_{ijk}(X)\xi^i\xi^j\xi^k \right)\left( DX^l - \xi^l\right) \right. \\
& \quad \qquad \left. - D\xi^i \left(D\Theta_i - p_i + \frac{1}{2} H_{ijk}(X)\xi^j\xi^k\right) \right) \\
&= \int_{T\left[1\right]N} \mu\, \frac{1}{6}\partial_l H_{ijk}(X)\xi^l\xi^i\xi^j\xi^k \\
&= 0.
\end{split}
\end{displaymath}
In the second to last line, we used Stokes' theorem to eliminate $D$-exact terms since $\partial N=\emptyset$, and the last equality follows from the fact that $dH=0$.
\section{Dualities in the BV Formalism} \label{BVdualities}
Two quantum field theories are called dual to each other if they describe equivalent physics, or equivalent topological invariants in the case of topological field theories, even though they are seemingly different. The BV formalism provides two natural ways to obtain dual theories, that we will describe here. The first one involves effective field theories, derived through a process called ``BV pushforward'', and the second one involves BV morphisms. Of course, one can combine BV pushforwards with BV morphisms to obtain a third composite way.
\subsection{BV Pushforwards} \label{effectivetheories}
The procedure to construct effective actions in the BV formalism was first suggested by Losev \cite{LosevEffective} and later used in \cite{effectiveBV}.

Suppose that the space of fields admits a splitting
\begin{displaymath}
\mathcal{F}=\mathcal{F}_{\mathrm{IR}} \oplus \mathcal{F}_{\mathrm{UV}}
\end{displaymath}
into infrared and ultraviolet degrees of freedom, compatible with a decomposition of the BV structure
\begin{displaymath}
\Omega = \Omega_{\mathrm{IR}} + \Omega_{\mathrm{UV}}
\end{displaymath}
in the sense that $\Omega_{\mathrm{IR}}$ is a BV structure on $\mathcal{F}_{\mathrm{IR}}$ and  $\Omega_{\mathrm{UV}}$ is one on $\mathcal{F}_{\mathrm{UV}}$, and that we have a solution $W$ of the QME on $\mathcal{F}$. Then we can integrate $e^{\frac{i}{\hbar}W}$ over a Lagrangian submanifold of the ultraviolet sector of the space of fields to find an effective BV action in the infrared sector,
\begin{equation}
e^{\frac{i}{\hbar}W_{\mathrm{eff}}} = \int_{\mathcal{L}_{\mathrm{UV}}\subset\mathcal{F}_{\mathrm{UV}}} \sqrt{\mu}_{\mathcal{L}_{\mathrm{UV}}} \, e^{\frac{i}{\hbar}W}.
\end{equation}
One can show that the effective action $W_{\mathrm{eff}}$ satisfies the QME associated to $\mathcal{F}_{\mathrm{IR}}$. In the physics language, one says that the ultraviolet degrees of freedom have been integrated out. In mathematics, one also talks about a ``BV pushforward'' by the projection map
$\rho_{\mathrm{UV}}: \mathcal{F} \rightarrow \mathcal{F}_{\mathrm{IR}}$
onto the infrared sector of the space of fields,
\begin{equation}
e^{\frac{i}{\hbar}W_{\mathrm{eff}}} = \rho_{\mathrm{UV}\ast} ( e^{\frac{i}{\hbar}W}).
\end{equation}
One step farther, we can pick a Lagrangian submanifold $\mathcal{L}_{\mathrm{IR}}$ of the infrared sector $\mathcal{F}_{\mathrm{IR}}$ and perform the functional integration of $e^{\frac{i}{\hbar}W_{\mathrm{eff}}}$ thereon to compute the partition function of the full model. This integration can also be represented by a pushforward map $\rho_{\mathrm{IR}\ast}$,
\begin{displaymath}
Z = \rho_{\mathrm{IR}\ast}( e^{\frac{i}{\hbar}W_{\mathrm{eff}}} ) = \rho_{\mathrm{IR}\ast} \circ \rho_{\mathrm{UV}\ast} ( e^{\frac{i}{\hbar}W}).
\end{displaymath}
This last step corresponds to the computation of the partition function on the Lagrangian submanifold $\mathcal{L}_{\mathrm{IR}} \times \mathcal{L}_{\mathrm{UV}}$ of the full space of fields $\mathcal{F}$. As a consequence of the Batalin-Vilkovisky theorem, the value of $Z$ does not depend on the particular splitting $\mathcal{F}=\mathcal{F}_{\mathrm{IR}} \oplus \mathcal{F}_{\mathrm{UV}}$, and a different choice $\mathcal{F}=\mathcal{F}'_{\mathrm{IR}} \oplus \mathcal{F}'_{\mathrm{UV}}$ leads to the same result,
\begin{equation}
\begindc{\commdiag}[50]
\obj(1,2){$e^{\frac{i}{\hbar}W}$}
\obj(0,1){$e^{\frac{i}{\hbar}W_{eff}}$}
\obj(2,1){$e^{\frac{i}{\hbar}W'_{eff}}$}
\obj(1,0){$Z=Z'$}
\mor(1,2)(0,1){$\rho_{UV\ast}$}[-1,\aplicationarrow]
\mor(1,2)(2,1){$\rho'_{UV\ast}$}[+1,\aplicationarrow]
\mor(0,1)(1,0){$\rho_{IR\ast}$}[-1,\aplicationarrow]
\mor(2,1)(1,0){$\rho'_{IR\ast}$}[+1,\aplicationarrow]
\enddc
\end{equation}
The models with action $W_{\mathrm{eff}}$ and $W'_{\mathrm{eff}}$ hence have the same partition function.

We should also include observables, if we want to compare correlation functions of the two models. We mention this only for the sake of completeness, and we will not go into many details, as the examples we will be treating below do not involve observables, the topological information we are interested in being completely encoded in the action.

Starting with an observable $f\in\mathcal{O}_{\mathrm{quant}}$ of the BV model on $\mathcal{F}$, we can pushforward $f\,e^{\frac{i}{\hbar}W}$ by $\rho_{UV}$ and $\rho'_{UV}$ to find $f_{eff}e^{\frac{i}{\hbar}W_{eff}}$ and $f'_{eff}e^{\frac{i}{\hbar}W'_{eff}}$ with identical correlation functions $\langle f_{eff} \rangle_{W_{eff}}  = \langle f'_{eff} \rangle_{W'_{eff}}  = \langle f \rangle_W$ (the subscript denotes the action of the model in which the corresponding correlation function is to be computed), and thus construct the algebras of observables $\mathcal{O}_{\mathrm{quant},eff}$ and $\mathcal{O}'_{\mathrm{quant},eff}$. Now in order to obtain truly dual theories, we should choose splittings of $\mathcal{F}$ and take $\mathcal{O}_{\mathrm{quant}}$ to be only a subalgebra of all the possible observables of the whole model in such a way that the algebras $\mathcal{O}_{\mathrm{quant},eff}$ and $\mathcal{O}'_{\mathrm{quant},eff}$ are isomorphic.
\subsection{BV Morphisms} \label{BVMorphisms}
\subsubsection{Classical BV Morphisms}
Given two BV spaces of fields $(\mathcal{F},\Omega)$ and $(\mathcal{F}',\Omega')$, we call a map
\begin{displaymath}
\Phi: \mathcal{F} \rightarrow \mathcal{F}'
\end{displaymath}
a classical BV morphism if it is a symplectomorphism with respect to the BV structures $\Omega$ and $\Omega'$, namely if $\Phi$ is a diffeomorphism and
\begin{equation}
\Phi^\ast(\Omega') = \Omega,
\end{equation}
where $\Phi^\ast$ stands for the pullback by $\Phi$ of differential forms. If we know a classical BV action $S'$, solution of the classical master equation on $\mathcal{F}'$, $\left\lbrace S',S' \right\rbrace_{\Omega'}=0$, we immediately get a solution of the classical master equation on $\mathcal{F}$ by pulling it back with $\Phi$, because
\begin{displaymath}
\left\lbrace \Phi^\ast(S'), \Phi^\ast(S') \right\rbrace_{\Phi^\ast(\Omega')} = \Phi^\ast\left( \left\lbrace S',S' \right\rbrace_{\Omega'} \right) = 0,
\end{displaymath}
and we obtain a morphism of classical BV manifolds,
\begin{displaymath}
\Phi: (\mathcal{F},\Omega,S) \rightarrow (\mathcal{F}',\Omega',S').
\end{displaymath}
For the same reason, a classical observable $f' \in \mathcal{O}'_{\mathrm{clas}}$ on $\mathcal{F}'$ that satisfies $\left\lbrace S',f' \right\rbrace_{\Omega'}=0$ can be pulled back by $\Phi$ to a function on $\mathcal{F}$ that automatically satisfies $\left\lbrace \Phi^\ast(S'), \Phi^\ast(f') \right\rbrace_{\Phi^\ast(\Omega')} = 0$.

The two actions $S$ and $S'$ then describe similar dynamics and symmetries and isomorphic algebras of classical observables. In the case of topological field theories, this formalism can encode the diffeomorphism invariance, and $\Phi$ would be an automorphism of $\mathcal{F}$, that could be continuously transformed into the identity. Of greater interest is the situation where $\mathcal{F}$ and $\mathcal{F}'$ are different, or when $\Phi$ cannot be cast into a continuous family of automorphisms of $\mathcal{F}$ of the identity. In these cases, we say that the models related by $\Phi$ are classically dual to each other. We will give examples farther.

When the duality arises out of the mathematical structures on the target space of a sigma model, we speak of target space duality, or shorter T-duality.
\subsubsection{Quantum BV Morphisms}
One step farther, if $\mathcal{L} \subset \mathcal{F}$ is a Lagrangian submanifold with respect to the symplectic structure $\Omega = \Phi^\ast(\Omega')$, then $\Phi(\mathcal{L}) \subset \Phi(\mathcal{F}) = \mathcal{F}'$ is also Lagrangian, with respect to $\Omega'$, so this classical duality might be extended to the quantum level.

A quantum BV morphism
\begin{displaymath}
\hat{\Phi}: (\mathcal{F},\Omega,\mu,W) \rightarrow (\mathcal{F}',\Omega',\mu',W')
\end{displaymath}
between two quantum BV manifolds is defined as a formal power series of maps $\hat{\Phi}\in\mathrm{Map}(\mathcal{F},\mathcal{F}')\left[\left[ \hbar \right]\right]$ such that $\hat{\Phi}^\ast(\Omega')=\Omega$, $\hat{\Phi}_\ast(\mu) = \mu'$ and $\hat{\Phi}^\ast(W') = W$.

In this case, the partition function (as well as all other correlation functions) is left invariant by $\hat{\Phi}$,
\begin{equation}
\begin{split}
Z &= \int_{\mathcal{L} \subset \mathcal{F}} \sqrt{\mu}_{\mathcal{L}} e^{\frac{i}{\hbar}W} =\int_{\mathcal{L} \subset \mathcal{F}} \sqrt{\mu}_{\mathcal{L}} e^{\frac{i}{\hbar}\hat{\Phi}^\ast(W')}
\\
& \qquad \quad =\int_{\hat{\Phi}(\mathcal{L}) \subset \hat{\Phi}(\mathcal{F})} \sqrt{\hat{\Phi}_\ast(\mu)}_{\hat{\Phi}(\mathcal{L})} e^{\frac{i}{\hbar}W'}= Z'.
\end{split}
\end{equation}

Unfortunately, quantum duality between $W$ and $W'$ does not necessarily follow from classical duality of their tree-level part, $W\vert_{\hbar=0} = \Phi^\ast(W'\vert_{\hbar=0})$, since most of the time a regularization scheme enters the game. Moreover, $\mu$, $\Omega$ and their duals might need to receive quantum corrections in $\hbar$. Consequently, quantum duality should always be checked independently from classical duality, which is possible order by order in $\hbar$.

Nevertheless, in what follows we will treat topological models with the property that the BV Laplacian can be regularized in such a way that the classical BV action is BV harmonic and thus already satisfies the quantum master equation, so that quantum duality actually follows from classical duality with $\hat{\Phi}= \Phi$.
\subsubsection{Target Space Duality} \label{atargetspaceduality}
With the AKSZ construction, we see that a trick to build BV morphisms is to find symplectomorphisms of target spaces of AKSZ models,
\begin{displaymath}
\Phi: (Y,\omega_Y) \rightarrow (Y',\omega_{Y'}).
\end{displaymath}
Such a symplectomorphism can be lifted to a BV morphism between the two AKSZ spaces of fields,
\begin{displaymath}
\Phi: \mathrm{Map}(T\left[1\right] N, Y) \rightarrow \mathrm{Map}(T\left[1\right] N, Y').
\end{displaymath}
Since the duality between the two resulting BV theories comes from a symplectomorphism of their target spaces, we call this duality target space duality, or T-duality. We will see a concrete example in section \ref{TdualityCSM}.
\subsection{Combination of BV Morphisms and BV Pushforwards} \label{combiBVmorphBVpf}
In a last step, we can naturally combine BV pushforwards with BV morphisms to construct dual BV effective theories. We start with a quantum BV morphism
\begin{displaymath}
\hat{\Phi}: (\mathcal{F},\Omega,W) \rightarrow (\mathcal{F}',\Omega',W')
\end{displaymath}
Essentially we get the same example of quantum duality as in the previous section, in particular the partition functions $Z$ and $Z'$ are equal. On each side of this BV morphism $\hat{\Phi}$, we choose a splitting of the space of fields into infrared and ultraviolet sectors, $\mathcal{F}=\mathcal{F}_{\mathrm{IR}} \oplus \mathcal{F}_{\mathrm{UV}}$ and $\mathcal{F}=\mathcal{F}'_{\mathrm{IR}} \oplus \mathcal{F}'_{\mathrm{UV}}$. Through respective BV pushforwards onto the infrared sectors, we obtain effective theories that admit the same partition function,
\begin{equation} \label{diagramBVmorphBVpf}
\begindc{\commdiag}[50]
\obj(0,2){$e^{\frac{i}{\hbar}W}$}
\obj(2,2){$\quad e^{\frac{i}{\hbar}W'}$}
\obj(0,1){$e^{\frac{i}{\hbar}W_{eff}}$}
\obj(2,1){$e^{\frac{i}{\hbar}W'_{eff}}$}
\obj(1,0){$Z=Z'.$}
\mor(2,2)(0,2){$\hat{\Phi}^\ast$}[-1,\aplicationarrow]
\mor(0,2)(0,1){$\rho_{UV\ast}$}[-1,\aplicationarrow]
\mor(2,2)(2,1){$\rho'_{UV\ast}$}[+1,\aplicationarrow]
\mor(0,1)(1,0){$\rho_{IR\ast}$}[-1,\aplicationarrow]
\mor(2,1)(1,0){$\rho'_{IR\ast}$}[+1,\aplicationarrow]
\enddc
\end{equation}
Note that a natural choice of UV sectors makes use of the BV morphism,
\begin{displaymath}
\mathcal{F}'_{\mathrm{UV}} = \hat{\Phi}(\mathcal{F}_{\mathrm{UV}}).
\end{displaymath}
This ensures that the effective theories are truly dual to each other, in particular we obtain isomorphic algebras of observables.
\section{T-duality and Courant Sigma Models} \label{TdualityCSM}
One of the simplest examples of duality expressed as a BV morphism involves the Courant sigma model based on a Courant algebroid of the form $(TE\oplus T^\ast E)/S^1$, where $E$ is a principal circle bundle, and actually reproduces the Courant algebroid isomorphism first discussed by Cavalcanti and Gualtieri \cite{TdualityCourant}.
\subsection{The Courant Sigma Model Based on $(TE\oplus T^\ast E)/S^1$}
We start with a principal circle bundle
\begin{displaymath}
\begin{CD}
S^1 @>>> E\\
@. @VV\rho V\\
@. M
\end{CD}
\end{displaymath}
over a manifold $M$, supporting a closed three-form $H\in\Omega^3_{\mathrm{closed}}(E)$ and equipped with a connection $\mathcal{A} \in \Omega^1(E)$ with curvature $F=d\mathcal{A}$. Evidently, we can use $H$ to define a Courant bracket on $TE\oplus T^\ast E$ and make it an exact Courant algebroid over $E$, as described in section \ref{CSM}. Now, if $H$ is $S^1$-invariant, we can actually restrict the structures defining the Courant algebroid (i.e. scalar product, Courant bracket and anchor map) to $S^1$-invariant sections of $TE$ and $T^\ast E$, and thus obtain a Courant algebroid over $M$ with total space $(TE\oplus T^\ast E)/S^1$, which is no longer exact.

Once we get the BV structure and BV action of the associated Courant sigma model, it will be more or less clear how to define a BV morphism to a dual Courant sigma model, but to reach this goal, we first need explicit expressions for the three structures defining a Courant algebroid.

A couple of tricks simplify this task. First, since $H$ is $S^1$-invariant, we can use the connection $\mathcal{A}$ to express its component along the fibers of $E$ and decompose it into two terms
\begin{displaymath}
H = H_{(3)} + \mathcal{A} \wedge H_{(2)},
\end{displaymath}
where $H_{(3)} \in \Omega^3(M)$ and $H_{(2)} \in \Omega^2(M)$ are two basic forms.

Second, we notice that the involved quotient bundles can be decomposed\footnote{This can be shown with an Atiyah exact sequence.} as
\begin{equation} \label{quotientedbundles}
TE/S^1 \cong TM \oplus \langle\partial_{\mathcal{A}}\rangle, \quad T^\ast E/S^1 \cong T^\ast M \oplus \langle\mathcal{A}\rangle,
\end{equation}
where $\partial_{\mathcal{A}}$ is an invariant period-1 generator of the circle action. Under this splitting, sections of $(TE\oplus T^\ast E)/S^1$ can be identified with the following expressions,
\begin{displaymath}
\begin{array}{rcl}
\mathcal{X} &=& X + f\,\partial_{\mathcal{A}} + \alpha + s\,\mathcal{A}, \\
\mathcal{Y} &=& Y + g\,\partial_{\mathcal{A}} + \beta + t\,\mathcal{A},
\end{array}
\end{displaymath}
where $X$ and $Y$ are vector fields on $M$, $\alpha$ and $\beta$ one-forms on $M$ and $f$, $g$, $s$ and $t$ real-valued functions on $M$.

We are now in a position to give an explicit expression of their scalar product,
\begin{equation} \label{invscapro}
\langle\mathcal{X},\mathcal{Y}\rangle = \frac{1}{2}(\alpha(Y)+\beta(X)+s\,g+t\,f),
\end{equation}
as well as of their Courant bracket \cite{Kao},
\begin{equation} \label{invCourantbracket}
\begin{split}
\left[\mathcal{X},\mathcal{Y}\right] =& \left[X,Y\right] + \left( X(g) - Y(f) + \imath_X \imath_Y F \right) \partial_{\mathcal{A}} \\
& +  \mathcal{L}_X \beta - \mathcal{L}_Y \alpha + t \imath_X F - s \imath_Y F - \frac{1}{2} \left(\imath_X \beta - \imath_Y \alpha \right)  \\
&+  \frac{1}{2} \left( df \, t + g ds - f dt - dg \, s \right) + \imath_X \imath_Y H_{(3)} + g \imath_X H_{(2)} - f \imath_Y H_{(2)} \\
&+ \left( X(t) - Y(s) + \imath_X \imath_Y H_{(2)} \right) \mathcal{A},
\end{split}
\end{equation}
and finally of the anchor map,
\begin{equation} \label{invanchor}
\rho(\mathcal{X})=X.
\end{equation}

Knowing the constituents of this Courant algebroid, we may construct a Courant sigma model for $(TE\oplus T^\ast E)/S^1$. The procedure only requires a small adaptation from the sigma model associated to an exact Courant algebroid. The space of fields is the mapping space between the odd tangent bundle of a three-dimensional manifold $N$ and the degree 2 cotangent bundle of the quotiented degree 1 cotangent bundle of the circle bundle $E$, namely
\begin{displaymath}
\mathcal{F} = \mathrm{Map}\left( T\left[1\right]N,T^\ast\left[2\right](T^\ast\left[1\right]E/S^1) \right).
\end{displaymath}
We can use the decomposition (\ref{quotientedbundles}) to identify this space of fields with
\begin{equation} \label{FinvCSM}
\mathcal{F} \cong  \mathrm{Map}\left( T\left[1\right]N,T^\ast\left[2\right]M \oplus T\left[1\right]M \oplus T^\ast\left[1\right]M \oplus \langle \partial_{\mathcal{A}}\rangle\left[1\right] \oplus \langle \mathcal{A} \rangle\left[1\right] \right).
\end{equation}
Like in the example of the Courant sigma model based on an exact Courant algebroid, we will work with superfields, namely a base map $X \in \mathrm{Map}( T\left[ 1\right] N, M)$, that we complete with fiber maps $p$, $\xi$ and $\Theta$ such that
\begin{displaymath}
\begin{array}{rcl}
(X,p) &\in & \mathrm{Map}( T\left[ 1\right] N, T^\ast\left[ 2\right] M), \\
(X,\xi) &\in & \mathrm{Map}( T\left[ 1\right] N, T\left[ 1\right] M), \\
(X, \Theta) &\in & \mathrm{Map}( T\left[ 1\right] N, T^\ast\left[ 1\right] M),
\end{array}
\end{displaymath}
and two odd functions $\phi,\psi\in\mathrm{Fun}(T\left[ 1\right] N,\mathbb{R}\left[ 1\right] )$ that we can combine with $\partial_{\mathcal{A}}$ and $\mathcal{A}$ respectively to obtain superfields associated to the last two components of the target space (\ref{FinvCSM}). In effect, we may identify the space of fields with the mapping space
\begin{displaymath}
\mathcal{F} \cong  \mathrm{Map}\left( T\left[1\right]N,T^\ast\left[2\right]M \oplus T\left[1\right]M \oplus T^\ast\left[1\right]M \oplus \mathbb{R}\left[1\right] \oplus \mathbb{R}\left[1\right] \right),
\end{displaymath}
with coordinate-fields $(X,p,\xi,\Theta,\phi,\psi)$.

The target space being a cotangent bundle, the space of fields carries the canonical BV structure
\begin{equation} \label{CSMBVstrE}
\Omega = \int_{T\left[1\right]N} \mu\, \left( \delta p_i \, \delta X^i - \delta\xi^i \, \delta\Theta_i - \delta \phi \,\delta\psi \right).
\end{equation}
The AKSZ action can be constructed by using the Courant bracket (\ref{invCourantbracket}) and anchor map (\ref{invanchor}) in the formula (\ref{AKSZaction}),
\begin{equation} \label{CSMactionE}
\begin{split}
S = \int_{T\left[1\right]N}\mu & \, \left( p_i \, DX^i + \frac{1}{2}\xi^i \, D\Theta_i + \frac{1}{2}\Theta_i \, D\xi^i + \frac{1}{2}\phi \,D\psi + \frac{1}{2}\psi \, D\phi \right. \\
& \left. -p_i\xi^i + \psi\,\frac{1}{2}F_{ij}\xi^i\xi^j + \frac{1}{6}H_{(3)ijk}\xi^i\xi^j\xi^k + \phi\,\frac{1}{2}H_{(2)ij}\xi^i\xi^j \right).
\end{split}
\end{equation}
In a similar way as the calculation ran in section \ref{AKSZconstr} for exact Courant algebroids, one can check that the classical master equation
\begin{displaymath}
\begin{split}
\frac{1}{2}\left\lbrace S,S\right\rbrace = \int_{T\left[1\right]N}  \mu \, & \left( - \psi\,\frac{1}{2}\partial_l F_{ij}\xi^l\xi^i\xi^j - \phi\,\frac{1}{2}\partial_l H_{(2)ij}\xi^l\xi^i\xi^j \right. \\
& \left. + \frac{1}{6}\partial_l H_{(3)ijk}\xi^l\xi^i\xi^j\xi^k -\frac{1}{2}F_{ij}\xi^i\xi^j\frac{1}{2}H_{(2)kl}\xi^k\xi^l \right) =0
\end{split}
\end{displaymath}
is satisfied provided
\begin{equation}
\begin{array}{rcl}
dF &=& 0, \\
dH_{(2)} &=& 0, \\
dH_{(3)} - F\wedge H_{(2)} &=& 0.
\end{array}
\end{equation}
The first condition follows from the fact that the curvature of a connection on a principal circle bundle is closed, and the other two from the fact that the twist $H$ is also closed.
\subsection{A T-duality BV Morphism}
If we take a closer look at the BV structure (\ref{CSMBVstrE}), we see that it is invariant under the exchange $\phi \leftrightarrow \psi$. The first obvious step in defining a BV morphism $\Phi:\mathcal{F} \rightarrow \hat{\mathcal{F}}$ would therefore be to require that the dual space of fields $\hat{\mathcal{F}}$ admits the same identification as the original space of fields $\mathcal{F}$,
\begin{equation} \label{dualtargetspaceisom}
\hat{\mathcal{F}} \simeq  \mathrm{Map}\left( T\left[1\right]N,T^\ast\left[2\right]M \oplus T\left[1\right]M \oplus T^\ast\left[1\right]M \oplus \mathbb{R}\left[1\right] \oplus \mathbb{R}\left[1\right] \right)
\end{equation}
with coordinate-fields $(X,p,\xi,\Theta,\hat{\phi},\hat{\psi})$, then set
\begin{displaymath}
\Phi(\phi) = \hat{\phi}=\psi \quad \mathrm{and} \quad \Phi(\psi) = \hat{\psi}=\phi,
\end{displaymath}
and ask that it leaves the other fields invariant,
\begin{displaymath}
\Phi \vert_{\mathrm{Map}\left( T\left[1\right]N,T^\ast\left[2\right](T^\ast\left[1\right]M) \right)} = \mathrm{Id}.
\end{displaymath}
The dual space of fields $\hat{\mathcal{F}}$ carries the natural BV structure
\begin{equation}
\hat{\Omega} = \int_{T\left[1\right]N} \mu\, \left( \delta p_i \, \delta X^i - \delta\xi^i \, \delta\Theta_i - \delta \hat{\phi} \,\delta \hat{\psi} \right),
\end{equation}
and $\Phi$ is automatically a BV morphism, $\Phi^\ast(\hat{\Omega}) = \Omega$.

In order to determine the dual BV theory, we still need to find an action $\hat{S}$ on $\hat{\mathcal{F}}$ such that $\Phi^\ast(\hat{S}) = S$. A functional of the coordinate fields of $\hat{\mathcal{F}}$ that satisfies this requirement is
\begin{equation}
\begin{split}
\hat{S} = \int_{T\left[1\right]N}\mu & \, \left( p_i \, DX^i + \frac{1}{2}\xi^i \, D\Theta_i + \frac{1}{2}\Theta_i \, D\xi^i + \frac{1}{2}\hat{\phi} \,D\hat{\psi} + \frac{1}{2}\hat{\psi} \, D\hat\phi \right. \\
& \left. -p_i\xi^i + \hat{\psi}\,\frac{1}{2}\hat{F}_{ij}\xi^i\xi^j + \frac{1}{6}\hat{H}_{(3)ijk}\xi^i\xi^j\xi^k + \hat{\phi}\,\frac{1}{2}\hat{H}_{(2)ij}\xi^i\xi^j \right),
\end{split}
\end{equation}
provided $\Phi$ acts also on the background fields (by `background fields', we understand fields defined on the target space of the model),
\begin{displaymath}
F \mapsto \Phi(F) = \hat{F}, \quad H_{(2)} \mapsto \Phi(H_{(2)}) = \hat{H}_{(2)} \quad H_{(3)} \mapsto \Phi(H_{(3)}) = \hat{H}_{(3)},
\end{displaymath}
such that
\begin{equation} \label{TdualityBackgroundFieldsCSM}
\hat{F} = H_{(2)}, \quad \hat{H}_{(2)} = F, \quad \hat{H}_{(3)} = H_{(3)}.
\end{equation}
In other words, the roles of the curvature $F$ and the component $H_{(2)}$ of the twist $H$ have been exchanged.  If the twist $H$ is chosen in such a way that $H_{(2)} = \hat{F}$ has integral periods (which is the case of the ones relevant to physics, due to the Wess-Zumino consistency condition), it can be related to the first Chern class of a principal circle bundle $\hat{E}$ with connection $\hat{\mathcal{A}}$ satisfying $\hat{F} = d\hat{\mathcal{A}}$. Then one can see that the dual action $\hat{S}$ describes a Courant sigma model for the Courant algebroid $(T\hat{E} \oplus T^\ast\hat{E})/S^1$ twisted by
\begin{displaymath}
\hat{H} = \hat{H}_{(3)} + \hat{H}_{(2)} \wedge \hat{\mathcal{A}} = H_{(3)} + F \wedge  \hat{\mathcal{A}} .
\end{displaymath}
The target space of the corresponding AKSZ construction
\begin{displaymath}
\hat{\mathcal{F}} = \mathrm{Map}\left( T\left[1\right]N,T^\ast\left[2\right](T^\ast\left[1\right]\hat{E}/S^1) \right)
\end{displaymath}
can be decomposed in a similar way as the original model,
\begin{displaymath}
\hat{\mathcal{F}}  \simeq  \mathrm{Map}\left( T\left[1\right]N,T^\ast\left[2\right]M \oplus T\left[1\right]M \oplus T^\ast\left[1\right]M \oplus \langle \partial_{\hat{\mathcal{A}}}\rangle\left[1\right] \oplus \langle \hat{\mathcal{A}} \rangle\left[1\right] \right),
\end{displaymath}
which allows us to describe the geometric structure of $\Phi$,
\begin{displaymath}
\begin{array}{rcl}
\Phi( \mathrm{Map}\left( T\left[1\right]N,\langle \partial_{\mathcal{A}}\rangle\left[1\right] \right) ) &=& \mathrm{Map} ( T\left[1\right]N, \langle \hat{\mathcal{A}} \rangle\left[1\right] ), \\
\Phi( \mathrm{Map}\left( T\left[1\right]N,\langle \mathcal{A} \rangle\left[1\right] \right) ) &=& \mathrm{Map}\left( T\left[1\right]N, \langle \partial_{\hat{\mathcal{A}}}\rangle\left[1\right] \right),
\end{array}
\end{displaymath}
and which ensures that the identification (\ref{dualtargetspaceisom}) is valid.

So if we assume that both spaces of fields are identified with the same model space of fields,
\begin{displaymath}
\mathcal{F}  \simeq \mathrm{Map}\left( T\left[1\right]N,T^\ast\left[2\right]M \oplus T\left[1\right]M \oplus T^\ast\left[1\right]M \oplus \mathbb{R}\left[1\right] \oplus \mathbb{R}\left[1\right] \right) \simeq \hat{\mathcal{F}},
\end{displaymath}
we can interpret the BV structures $\Omega$ and $\hat{\Omega}$ and BV actions $S$ and $\hat{S}$ as functionals on this model space of fields, on which we even have the equalities $\hat{\Omega}=\Omega$ and $\hat{S} = S$. These ensure that the Courant algebroids $(TE\oplus T^\ast E)/S^1$ and $(T\hat{E}\oplus T^\ast \hat{E})/S^1$ twisted by $H$ and $\hat{H}$ respectively are actually isomorphic, as the Courant algebroid structures are encoded in the associated Courant sigma model actions \cite{AKSZCSM}.

This isomorphism is the same as the one derived by Cavalcanti and Gualtieri in \cite{TdualityCourant}  through arguments solely based on geometrical considerations inspired by T-duality in string theory. Our field theoretic approach, on the other hand, follows the same spirit as the derivation of the Buscher rules from a duality of sigma models. We now give a short review of T-duality in physics and geometry to illustrate the difference between the two approaches and to motivate our next example of duality in the BV formalism, based this time on BV pushforwards combined with BV morphisms.
\section{T-duality in Physics and Geometry} \label{Tduality}
In this review, we focus on the results that can be expressed as examples of dual BV theories. References for standard material are provided.
\subsection{Periodic Scalar Field}
The simplest example of T-duality (see \cite{HoriVafa} for details) arises when one considers the bosonic theory of a single scalar field $\phi$ of periodicity $2\pi$ on a worldsheet $\Sigma$, with action
\begin{displaymath}
S_\phi = \frac{R^2}{2} \int_\Sigma d\phi \wedge \ast d\phi,
\end{displaymath}
where the Hodge star operator is denoted by $\ast$. We assume the worldsheet metric $g_\Sigma$ to be of Lorentzian signature so that $\ast$ squares to one when applied on one-forms. We recognize the action of a sigma model whose target space is a circle of radius $R$. We can introduce an auxiliary one-form field $\eta$ ton construct the first order action
\begin{displaymath}
S' = \frac{1}{2R^2}\int_\Sigma \eta \wedge \ast \eta + \int_\Sigma \eta \wedge d\phi.
\end{displaymath}
If we complete the square to integrate out $\eta$, we recover the original action.

On the other hand, if we first integrate over $\phi$, it imposes the constraint $d\eta=0$, which can be shown to be equivalent \cite{HoriVafa} to
\begin{displaymath}
\eta = d\vartheta
\end{displaymath}
for some dual periodic scalar $\vartheta$, also of period $2\pi$. Inserting this condition into the extended action $S'$ leads to the T-dual action
\begin{displaymath}
S_\vartheta = \frac{1}{2R^2} \int_\Sigma d\vartheta \wedge \ast d\vartheta,
\end{displaymath}
another sigma model with a circle for its target space, but with radius $1/R$. One can also find a direct relation between $\phi$ and $\theta$,
\begin{equation} \label{exchangemomentumwinding}
Rd\phi = \frac{1}{R}\ast d\vartheta.
\end{equation}
Since $Rd\phi$ and $R\ast d\phi$ are the conserved currents of the theory with action $S_\phi$ that count the momentum and the winding number respectively, the relation (\ref{exchangemomentumwinding}) means that T-duality not only transforms the radius $R\leftrightarrow 1/R$, but also exchanges the momentum and the winding number around the circle \cite{HoriVafa}.
\subsection{Principal Circle Bundles, Buscher Rules, Curvature and H-flux}
The sigma models of the previous section can be interpreted as string theories. However, a consistent string theory cannot admit a one-dimensional target space such as $S^1$. It has to be completed with a nine-dimensional manifold $M$ to form a ten-dimensional target space $\mathcal{M} = S^1\times M$ compatible with superstring theories. More generally, one can also consider the ten-dimensional total space $E$ of some principal circle bundle over $M$,
\begin{displaymath}
\begin{CD}
S^1 @>>> E\\
@. @VV\rho V\\
@. M.
\end{CD}
\end{displaymath}
In this case, the simple exchange $R\leftrightarrow 1/R$ for a single scalar field $\phi$ and its T-dual $\vartheta$ is replaced by the more complicated Buscher rules \cite{Buscherrules1} \cite{Buscherrules2}.

The starting point for their derivation is the Polyakov action of the string sigma model with background fields. In this model, strings are parametrized by maps $X:\Sigma\rightarrow \mathcal{M}$ from a two-dimensional worldsheet $\Sigma$ to a space-time manifold $\mathcal{M}$. If $(x^I)_{I=1}^N$ is a set of coordinates on a patch $U$ of $\mathcal{M}$, we can decompose $X$ in its components $X^I$ for convenience. The worldsheet $\Sigma$ supports a metric $g_\Sigma$ while the target space $\mathcal{M}$ comes equipped with a metric $G=G_{IJ}(x)dx^I\otimes dx^J$ and a B-field $B=\frac{1}{2}B_{IJ}(x)dx^I\wedge dx^J$ (in mathematical terms, the connection of a two-gerbe) with curvature $H=dB$, the H-flux, both corresponding to massless modes of the string spectrum. The dynamics is described by the action
\begin{equation} \label{stringsigmamodel}
S_{\mathrm{string}}=\int_\Sigma \frac{1}{2}G_{IJ}dX^I\wedge\ast dX^J + \frac{1}{2}B_{IJ}dX^I\wedge dX^J.
\end{equation}

It is easy to see how $S_\phi$ is a one-dimensional version of this action, with the single component of the metric corresponding to the compactification radius $R$.

To find the Buscher rules, one can thus consider $S_{\mathrm{string}}$ with a target space given by the total space $E$ of a principal circle bundle, and T-dualize along its fibers.

If $\mathcal{A}$ is a connection of $E$, we choose coordinates on $E$ such that it is written as $\mathcal{A}=dx^0 + A_i\,dx^i$, where $x^0$ is a coordinate along the fibers and the $x^i$'s are coordinates on the base manifold $M$. The part $A=A_i\, dx^i$ is sometimes called the gauge potential.

The connection $\mathcal{A}$ allows us to write a canonical invariant metric as well as a B-field on $E$,
\begin{equation} \label{metricBfieldonE}
\begin{array}{rcl}
G &=& \mathcal{A}\otimes\mathcal{A} + g_{ij} dx^i\otimes dx^j, \\
B &=& \hat{A}\wedge\mathcal{A} + \frac{1}{2}b_{ij} dx^i\wedge dx^j,
\end{array}
\end{equation}
and we define $b=\frac{1}{2}b_{ij} dx^i\wedge dx^j$.

The sigma model obtained when these structures are introduced in the action (\ref{stringsigmamodel}) possesses a global $U(1)$ symmetry $X^0\rightarrow X^0 + C$ that can be gauged \cite{Buscherrules1}, \cite{Buscherrules2} by introducing a $U(1)$ connection $\theta$ on $\Sigma$ and a Lagrange multiplier $\hat X_0$ to enforce flatness of $\theta$. If one integrates out this Lagrange multiplier, one retrieves the original model, but if one integrates out the connection $\theta$, one obtains a T-dual sigma model based on a T-dual circle bundle $\hat E$ with fiber coordinate $\hat x_0$ (defined via the Lagrange multiplier field $\hat{X}_0$) supporting background fields $\hat G$ and $\hat B$, related to the original ones through the Buscher rules \cite{Buscherrules1} \cite{Buscherrules2}.
%\begin{equation} \label{Buscher}
%\hat G_{00} = \dots
%\end{equation}

The topological content of these rules has been formalized by Bouwknegt, Evslin and Mathai \cite{Bouwknegt1} \cite{Bouwknegt2} \cite{Bouwknegt3}. Essentially, the potential $\hat A$ defined through the relation $B = b + \hat A\wedge\mathcal{A}$ turns out to enter the definition of a connection for the dual bundle, $\hat{\mathcal{A}}=d\hat{x}_0 + \hat{A}$, and the T-dual B-field is given by
\begin{equation} \label{TdualB}
\hat B = b + A\wedge d\hat{x}^0 = b + A\wedge \hat{\mathcal{A}} - A\wedge\hat{A}.
\end{equation}
From there on, it is easy to calculate the H-flux and its T-dual,
\begin{equation} \label{HhatH}
\begin{array}{rcl}
H &=& db - \hat{A}\wedge F + \hat{F} \wedge \mathcal{A}, \\
\hat H &=& db - \hat{A}\wedge F + F \wedge \hat{\mathcal{A}}.
\end{array}
\end{equation}

Note that out of (\ref{TdualB}), we find can also find the T-dual partner of $b$,
\begin{equation} \label{Tdualb}
\hat{b} = b - A\wedge\hat{A}.
\end{equation}
This relation can be symmetrized to obtain a (local) two-form
\begin{equation}
\hat{b} - \frac{1}{2} \hat{A} \wedge A = b - \frac{1}{2} A\wedge\hat{A} =: b_{\mathrm{inv}}
\end{equation}
which is invariant under T-duality. The relation (\ref{TdualB}) then corresponds to the exchange of the circle bundles $E$ and $\hat{E}$ in the formulas
\begin{equation} \label{bhatb}
b = b_{\mathrm{inv}} + \frac{1}{2} A\wedge\hat{A} \leftrightarrow \hat{b} = b_{\mathrm{inv}} + \frac{1}{2} \hat{A} \wedge A.
\end{equation}

The topological information of each of these sigma models is contained in the H-fluxes $H$ and $\hat{H}$ and the curvatures $F$ and $\hat{F}$ (we recall that the first Chern class $\left[F\right] \in H^2(M;\mathbb{Z})$ of the associated line bundle uniquely characterizes a circle bundle $E$). From the relations  (\ref{HhatH}) between these four differential forms, we can therefore extract the topological content of the Buscher rules and define geometric T-duality as follows: T-duality for principal circle bundles relates two pairs $(E,H)$ and $(\hat{E},\hat{H})$ of principal circle bundles $E$ and $\hat{E}$ over a mutual base-manifold $M$ and H-fluxes $\left[H\right]\in H^3(E;\mathbb{Z})$ and $\left[\hat{H}\right]\in H^3(\hat{E};\mathbb{Z})$. If $\mathcal{A}$ is a connection of $E$ with curvature $F$ and one assumes $H$ to be the $S^1$-invariant representative of $\left[H\right]$, and the same yields for $\hat{\mathcal{A}}$, $\hat{E}$, $\hat{F}$ and $\hat{H}$, one can decompose these fluxes as
\begin{displaymath}
H = H_{(3)} + \mathcal{A} \wedge H_{(2)} \ \mathrm{and} \ \hat{H} = \hat{H}_{(3)} + \hat{\mathcal{A}} \wedge \hat{H}_{(2)}.
\end{displaymath}
The pairs $(E,H)$ and $(\hat{E},\hat{H})$ are then called T-dual if \cite{Bouwknegt1} \cite{Bouwknegt2} \cite{Bouwknegt3}
\begin{equation} \label{topcontentBuscher}
F = \hat{H}_{(2)},\ \hat{F} = H_{(2)} \ \mathrm{and} \ H_{(3)} = \hat{H}_{(3)}.
\end{equation}
Note that these relations coincide with the action of our BV morphism on background fields of the Courant sigma model (\ref{TdualityBackgroundFieldsCSM}).

More formally, we can pullback the twists $H$ and $\hat{H}$ to the correspondence space $E \times_M \hat{E}$, namely the fiber product of the two bundles, by the dual projection maps $p: E \times_M \hat{E} \rightarrow E$ and $\hat{p}$ to compare them,
\begin{displaymath}
p^\ast(H) - \hat{p}^\ast(\hat{H}) = d ( \hat{p}^\ast(\hat{\mathcal{A}}) \wedge p^\ast(\mathcal{A}) ).
\end{displaymath}
This motivates the more general definition of T-duality that states that $(E,H)$ and $(\hat{E},\hat{H})$ are T-dual if there exists a non-degenerate two-form $\mathcal{B} \in \Omega( E \times_M \hat{E})$ on the correspondence space such that
\begin{displaymath}
p^\ast(H) - \hat{p}^\ast(\hat{H}) = d\mathcal{B}.
\end{displaymath}
This definition can be more easily extended to principal torus bundles and it can be justified by exact sequences in topology \cite{Bouwknegt3}, independently from the Buscher rules.
\subsection{Courant Algebroids}
This T-duality relation between pairs of principal circle bundles with H-flux can be used to define an isomorphism of complexes of $S^1$-invariant differential forms with twisted differential  \cite{Bouwknegt2},
\begin{displaymath}
\tau: (\Omega^\bullet_{S^1}(E), d_H) \rightarrow (\Omega^\bullet_{S^1}(\hat{E}), d_{\hat{H}}),
\end{displaymath}
where $d_H = d + H \wedge \cdot$. Explicitly, if $\omega\in \Omega^\bullet_{S^1}(E)$, we can write it as $\omega = \omega' + \mathcal{A} \wedge \omega''$, and its image as
$\tau(\omega) = \hat\omega = \hat\omega' + \hat{\mathcal{A}} \wedge \hat\omega''$, with
$\hat\omega' = \omega''$ and $ \hat\omega'' = -\omega'$. The isomorphism of complexes of twisted differentials means that $d_H\omega = 0$ if and only if $d_{\hat{H}}\hat{\omega}=0$.

Gualtieri and Cavalcanti use this isomorphism as the starting point for the construction of an isomorphism of Courant algebroids \cite{TdualityCourant}, that actually coincides with the one underlying our BV morphism.

Their first observation is that the space $\Omega^\bullet_{S^1}(E)$ of invariant differential forms on $E$ has the structure of a Clifford module for the Courant algebroid $(TE\oplus T^\ast E)/S^1$ over $M$ that we already encountered in our construction of a BV morphism. The Clifford action of a section of the Courant algebroid on a differential form is defined as the addition of the contraction with the vector field part and the exterior multiplication with the differential form part. An isomorphism of Courant algebroids
\begin{displaymath}
\Phi: (TE\oplus T^\ast E)/S^1 \rightarrow (T\hat{E}\oplus T^\ast \hat{E})/S^1,
\end{displaymath}
with $\hat{E}$ being the dual circle bundle, can then be constructed in such a way that the map $\tau: \Omega^\bullet_{S^1}(E) \rightarrow \Omega^\bullet_{S^1}(\hat{E})$ is promoted to an isomorphism of Clifford modules, namely such that
\begin{displaymath}
\tau(v\cdot\rho) = \Phi(v) \cdot \tau(\rho)
\end{displaymath}
for any section $v\in\Gamma((TE\oplus T^\ast E)/S^1)$ and any invariant differential form $\rho\in \Omega^\bullet_{S^1}(E)$. The Clifford action is denoted by the dot.

Their construction follows geometric considerations which are inspired from a duality of field theories, yet it could be based on purely topological and geometrical arguments. Our construction somehow closes the gap by providing a field theoretic derivation of the same isomorphism. In a next step, it is tempting to try to express the topological content of the Buscher rules as a BV duality of two-dimensional topological sigma models.
\section{T-duality and Twisted Poisson Sigma Models} \label{TdualityPSM}
This time, we will have to combine a BV morphism with dual BV pushforwards, a method explained in section \ref{combiBVmorphBVpf}. The BV morphism will involve a model constructed on the same correspondence space where we compared $H$ and $\hat{H}$. The pushforwards will be needed to go down to the individuel principal circle bundles, they will somehow correspond to the path integrations that led to the Buscher rules.
\subsection{A Sigma Model for the Topological Sector of a String with Background Fields}
In order to find a topological sigma model related to string theory and subject to T-duality transformations, it is natural to start with a WZ Poisson sigma model \cite{WZPSM}, whose action we recall is
\begin{equation}
S=\int_\Sigma \eta_i\wedge dX^i + \frac{1}{2}\pi^{ij}(X)\eta_i\wedge\eta_j + \int_N X^\ast(H).
\end{equation}
Here the $X^i$'s denote the coordinate components of a map $X\in\mathrm{Map}(\Sigma\rightarrow \mathcal{M})$, $\eta\in\Gamma(T^\ast\Sigma\otimes X^\ast (T^\ast \mathcal{M}))$ is a one form on the closed worldsheet $\Sigma$ with value in the pullback by $X$ of the cotangent bundle of the target space, $H\in\Omega^3_{\mathrm{closed}}(\mathcal{M})$ is the twist (or in the language of strings the H-flux) of the target space, and $N$ is a handlebody for $\Sigma = \partial N$. The Wess-Zumino consistency condition requires the twist to have integral period, $\left[ H \right] \in H^3(\mathcal{M};\mathbb{Z})$.

This choice of sigma model is motivated by the fact that when $\pi=0$, this model represents the topological sector of a sigma model for a string in background fields $G$ (metric on the target space $\mathcal{M}$) and $B$ (connection of a two-gerbe with curvature $dB=H$). Indeed, if we add a term $-\frac{1}{2}\int_\Sigma G^{ij}(X)\eta_i\wedge\ast\eta_j$ to the action, where $\ast$ denotes the Hodge star operator on the worldsheet $\Sigma$, we may integrate out the $\eta$ fields and recover the desired action.

The minimal BV extension of this simplified model with a zero Poisson structure requires to augment the classical space of fields spanned by $(X,\eta)$ with the algebra of symmetries shifted by one, that is the space of ghost fields $\beta\in\Gamma (X^\ast T^\ast\left[1\right] \mathcal{M})$, to form the BRST space of fields $\mathcal{F}_{\mathrm{BRST}}$ spanned by $(X,\eta,\beta)$, and then take its cotangent bundle shifted by minus one, $\mathcal{F}=T^\ast\left[-1\right]\mathcal{F}_{\mathrm{BRST}}$. This BV space of fields admits the canonical BV structure
\begin{equation}
\Omega = \int_\Sigma \delta\eta^{+i}\wedge\delta\eta_i + \delta X^+_i\,\delta X^i + \delta\beta^{+i}\,\delta\beta_i,
\end{equation}
where $\eta^+\in\Gamma(T^\ast\Sigma\otimes X^\ast T\left[-1\right]\mathcal{M})$, $X^+\in\Gamma(\bigwedge^2 T^\ast\Sigma\otimes X^\ast T^\ast\left[-1\right] \mathcal{M})$ and $\beta^+\in\Gamma(\bigwedge^2 T^\ast\Sigma\otimes X^\ast T\left[-2\right] \mathcal{M})$ are cotangent fiber coordinates of $\mathcal{F}$. The BV action is obtained through addition to the classical one of a term that encodes its symmetry under infinitesimal transformations $\delta_\epsilon\eta_i=d\epsilon_i$,
\begin{equation}
S=\int_\Sigma \eta_i\wedge dX^i - \eta^{+i}\wedge d\beta_i + \int_N X^\ast(H),
\end{equation}
and it remains to check that the classical master equation is indeed satisfied,
\begin{displaymath}
\begin{split}
\frac{1}{2}\left\lbrace S,S \right\rbrace &=\int_\Sigma \left( 
\frac{S \overleftarrow{\delta}}{\delta \eta_i} \frac{\overrightarrow{\delta}S}{\delta \eta^{+i}} +
\frac{S \overleftarrow{\delta}}{\delta X^i} \frac{\overrightarrow{\delta}S}{\delta X^+_i} +
\frac{S \overleftarrow{\delta}}{\delta \beta_i} \frac{\overrightarrow{\delta}S}{\delta \beta^{+i}} \right) \\
&= \int_\Sigma d X^i \wedge d \beta_i = -\int_\Sigma d(\beta_i dX^i) =0,
\end{split}
\end{displaymath}
as the integral of an exact form over a closed surface vanishes. Here functional left- and right-derivatives are defined in a similar way as in the case (\ref{functionalderivativessuperfields}) of superfields, the only difference being that one integrates here differential forms over a manifold instead of functions over a supermanifold with a canonical measure.

Note that the space of fields admits the structure of a mapping space \emph{\`a la} AKSZ, $\mathcal{F}=\mathrm{Map}(T\left[1\right]\Sigma,T^\ast\left[1\right] \mathcal{M})$, similar to the one of the regular Poisson sigma model, however the twist $H$ prevents the application of the full AKSZ procedure.

To investigate T-duality, we are interested in the situation where the target space is a principal circle bundle $E$ over $M$,
\begin{displaymath}
\begin{CD}
S^1 @>>> E\\
@. @VV\rho V\\
@. M.
\end{CD}
\end{displaymath}
We choose similar coordinates on $E$ as before, so that the connection is written as $\mathcal{A}=d x^0 + A$.

From covariant considerations, we would expect an action of the form
\begin{equation} \label{PSM_E}
S_E = \int_\Sigma \eta_i\wedge dX^i + \eta_0\wedge (dX^0 + X^\ast(A)) + \int_N X^\ast(H),
\end{equation}
where $dX^0 + X^\ast(A)$ is the pullback by the field $X$ of the connection $\mathcal{A}$ on $E$. If we add the term
\[
- \int_\Sigma \left( G^{ij}(X) \eta_i \wedge \ast \eta_j + G^{00}(X) \eta_0 \wedge \ast \eta_0 \right)
\]
to this topological action and integrate out the one-form fields $\eta$, we recover the string sigma model action with background fields given by the canonical metric and B-field (\ref{metricBfieldonE}) on the circle bundle $E$.

Furthermore, we want $H$ to be $S^1$-invariant, for reasons that will become clearer.

The price to pay for the introduction of the pullback of the connection $\mathcal{A}$ on $E$ is that the action is no longer invariant under transformations $\delta_{\epsilon_0}\eta_0 = d\epsilon_0$. It would pick up a term $\delta_{\epsilon_0}S_E=\int_\Sigma \epsilon_0\,F$ proportional to the curvature of the connection $\mathcal{A}$. One way to recover the symmetry would be to add a term $\int_\Sigma \hat X_0\,F=\int_\Sigma d\hat X_0 \wedge \mathcal{A}$ to the action and require that the new field $\hat X_0$ transforms as $\delta_{\epsilon_0}\hat X_0 = \epsilon_0$. Note that we have performed an integration by parts to make the dependence on the connection of $E$ obvious. The new symmetry is obviously abelian and one-dimensional, so it is natural to interpret $\hat X_0$ as induced by the fiber coordinate $\hat{x}_0$ of a dual bundle $\hat E$, to which we assign a connection $\hat{\mathcal{A}}=d\hat x_0 + \hat A$ with curvature $\hat F=d\hat{\mathcal{A}}$. Here dual means that the Lie algebras $\mathfrak{u}(1) \simeq \mathbb{R}$ and  $\hat{\mathfrak{u}}(1) \simeq \mathbb{R}$ of the fibers of $E$ and $\hat{E}$ are dual to each other, in the same way as for the circle bundles of section \ref{Tduality}.
\subsection{A BV Model on the Correspondence Space}
We are apparently building a model on the correspondence space $E\times_M\hat E$,
%= \left\lbrace (q,\hat{q}) \in E\times \hat{E} \ \vert \ \rho(q)=\hat{\rho}(\hat{q})\in M \right\rbrace,
the fiber product of both circle bundles $E$ and $\hat{E}$. It is therefore natural to gauge the $U(1)$ symmetry along fibers of $E$ by adding a connection one-form $\hat{\eta}^0$ for the pullback bundle $X^\ast(E)$ on $\Sigma$. We are now in possession of all the elements to write down a $U(1)\times U(1)$ invariant (classical) action on the correspondence space,
\begin{equation} \label{clactioncorrsp}
S_{E\times_M\hat E}^{\mathrm{cl}} = \int_\Sigma \eta_i\wedge dX^i + \left( \eta_0 + d\hat X_0 + X^\ast(\hat A) \right) \wedge \left( \hat{\eta}^0 + dX^0 + X^\ast(A) \right) + \int_N X^\ast(h).
\end{equation}
The three-form $h$ that enters the last term (a Wess-Zumino type term) is yet to be determined. The $U(1)\times U(1)$ invariance actually requires $h$ to be basic (i.e. $h\in\Omega^3_{\mathrm{closed}}(M)$).

Three geometrical structures of the target space $E\times_M\hat E$ enter the formulation of this model, namely the three-form $h$ and the two connections $\mathcal{A}$ and $\hat{\mathcal{A}}$. We will see how they behave under a T-duality transformation, but before we can study this example of a BV morphism, we obviously need to determine the BV formulation of the model.

Symmetry transformations are described by local parameters $\epsilon = (\epsilon_i,\epsilon_0,\hat{\epsilon}^0)$,
\begin{displaymath}
\begin{array}{lcl}
\delta_\epsilon \eta_i &=& d\epsilon_i, \\
\delta_\epsilon \eta_0 &=& d\epsilon_0, \\
\delta_\epsilon \hat{\eta}^0 &=& d\hat{\epsilon}^0, \\
\delta_\epsilon X^0 &=& \hat{\epsilon}^0, \\
\delta_\epsilon \hat X_0 &=& \epsilon_0.
\end{array}
\end{displaymath}
Again, to obtain the minimal BV formulation, we replace the gauge parameters with odd fields of ghost number 1, namely $\beta_i$, $\beta_0$ and $\hat{\beta}^0$, we assign an antifield to each field or ghost, we construct the canonical BV structure on the resulting space of fields $\mathcal{F}_{E\times_M\hat E}=\mathrm{Map}(T\left[1\right]\Sigma, T^\ast \left[1\right] (E \times_M \hat{E}))$,
\begin{equation} \label{CorrSpaceBVStr}
\begin{split}
\Omega_{E\times_M\hat E} = \int_\Sigma & \delta\eta^{+i}\wedge\delta\eta_i + \delta X^+_i\,\delta X^i + \delta\beta^{+i}\,\delta\beta_i \\
& + \delta\eta^{+0}\wedge\delta\eta_0 + \delta X^+_0\,\delta X^0 + \delta\beta^{+0}\,\delta\beta_0 \\
& + \delta\hat\eta^+_0\wedge\delta\hat\eta^0 + \delta \hat X^{+0}\,\delta \hat X_0 + \delta\hat\beta^+_0\,\delta\hat\beta^0,
\end{split}
\end{equation}
and finally we add corresponding terms that encode the symmetries to the action,
\begin{equation} \label{CorrSpaceBVAct}
S_{E\times_M\hat E} = S_{E\times_M\hat E}^{\mathrm{cl}} + \int_\Sigma - \eta^{+i}\wedge d\beta_i - \eta^{+0}\wedge d\beta_0 - \hat{\eta}^+_0 \wedge d\hat{\beta}^0 + X^+_0 \wedge \hat{\beta}^0 + \hat X^{+0}\wedge \beta_0.
\end{equation}

Note that even though the space of fields is similar to the usual mapping space of the AKSZ construction with its canonical BV structure, the BV action is not of the AKSZ type, like in the case of the twisted Poisson sigma model with trivial Poisson structure we considered above.

The BV action $S_{E\times_M\hat E}$ readily satisfies the quantum master equation, provided a suitable regularization of the BV Laplacian is adopted. It therefore corresponds to the action $W$ in the diagram (\ref{diagramBVmorphBVpf}).
\subsection{The Original Model as an Effective Theory}
We started from a sigma model on the circle bundle $E$ to build this action on the correspondence space. We thus expect to be able to re-derive the initial model as an effective action, as described in section \ref{effectivetheories}. This will correspond to the ultraviolet BV pushforward $\rho_{\mathrm{UV}\ast}$ in the diagram (\ref{diagramBVmorphBVpf}).

Obviously, to obtain an effective action on $E$, the ultraviolet sector of the space of fields needs to contain the components associated to $\hat E$, namely $\hat X_0$, $\hat{\eta}^0$ and $\hat{\beta}^0$. However, we saw that the symmetry associated to $\beta_0$ is also broken in the action (\ref{PSM_E}), consequently this ghost field should also belong to the UV sector. In summary, the ultraviolet sector $\mathcal{F}_{UV}$ is spanned by
\[
\left( \hat X_0, \hat{\eta}^0, \hat{\beta}^0, \beta_0, \hat X^{+0}, \hat{\eta}^+_0, \hat{\beta}^+_0, \beta^{+0} \right).
\]
We specify the Lagrangian subspace of $\mathcal{F}_{\mathrm{UV}}$ by setting $\beta_0=0$ and $\hat{\beta}^0=0$, which breaks the $U(1)\times U(1)$ symmetry of the total model, as well as $\hat X_0=0$ and $\hat{\eta}^0=0$, which selects only the fibers of $E$ in the model and removes the ones of the dual $\hat E$. Since the restriction of the action to this Lagrangian subspace of the ultraviolet sector does not depend on the other fields of this sector, the functional integration yields only trivial constants of no interest, and the resulting effective theory is described by the action
\begin{equation}
S^{\mathrm{eff}}_E =  \int_\Sigma \eta_i\wedge dX^i + \eta_0\wedge X^\ast(\mathcal{A}) + X^\ast(\hat A \wedge \mathcal{A}) + \eta^{+i}\wedge d\beta_i  + \int_N X^\ast(h).
\end{equation}
At first glance, the term $X^\ast(\hat A\wedge\mathcal{A})$ seems awkward, but it actually contributes to the WZ term. If we take its exterior derivative, we can write it as an integral over the handlebody $N$ and the connection $\mathcal{A}$ will bring an invariant contribution along the fibers of $E$ to a twist $H$ that would otherwise remain basic,
\begin{equation}
S^{\mathrm{eff}}_E =  \int_\Sigma \eta_i\wedge dX^i + \eta_0\wedge X^\ast(\mathcal{A}) + \eta^{+i}\wedge d\beta_i  + \int_N X^\ast(h-F\wedge\hat A + \hat F\wedge\mathcal{A}).
\end{equation}
We immediately see that the twist on the target space $E$ is given by
\begin{equation} \label{Hflux1}
H = h-F\wedge\hat A + \hat F\wedge\mathcal{A},
\end{equation}
and we understand why we had to assume that $H$ was $S^1$-invariant. Furthermore, by comparison with (\ref{HhatH}), we see that we can interpret $h$ as the curvature $h=db$ of the basic component of the B-field (\ref{metricBfieldonE}).
\subsection{A T-duality BV Morphism}
The next step is to determine the T-duality BV morphism $\Phi$, the one that sits at the top of the diagram (\ref{diagramBVmorphBVpf}), which will give us a dual action on the full space.

We saw that on two-dimensional models, T-duality involves the Hodge star operator. If we consider the term
\begin{displaymath}
\int_\Sigma  \left( \eta_0 + d\hat X_0 + X^\ast(\hat A) \right) \wedge \left( \hat{\eta}^0 + dX^0 + X^\ast(A) \right)
\end{displaymath}
of the action (\ref{clactioncorrsp}) of the sigma model on the correspondence space and apply on it the prescription (\ref{exchangemomentumwinding}) we found for T-duality\footnote{Note that in this topological model, the radius is normalized to $R=1$. The corresponding information is actually contained in the metric component $G_{00}$ that we left aside.}, we find the T-dual term
\begin{equation} \label{HodgeSwap}
\begin{split}
& \int_\Sigma  \ast\left( \eta_0 + d\hat X_0 + X^\ast(\hat A) \right) \wedge \ast\left( \hat{\eta}^0 + dX^0 + X^\ast(A) \right) \\
=& \int_\Sigma  \left( \hat{\eta}^0 + dX^0 + X^\ast(A) \right) \wedge \left( \eta_0 + d\hat X_0 + X^\ast(\hat A) \right),
\end{split}
\end{equation}
where we used the symmetry of the product $\cdot \wedge\ast \cdot$ of two one-forms and the involutivity of the Hodge star operator induced by a metric of Lorentzian signature when acting on one-forms. We see that T-duality can be interpreted as the exchange of the roles of the two circle bundles $E$ and $\hat{E}$, and that the Hodge star operator disappears from the formula, which is good when one considers topological field theories (of the Schwarz type).

We can use this first hint to start constructing a BV morphism $\Phi$, by requiring
\begin{displaymath}
\begin{array}{rclrcl}
\Phi(\eta_0) &=& \hat{\eta}^0, & \Phi(\hat{\eta}^0) &=& \eta_0, \\
\Phi(X^0) &=& \hat{X}_0, & \Phi(\hat{X}_0) &=& X^0, \\
\Phi(\beta_0) &=& \hat{\beta}^0, & \Phi(\hat{\beta}^0) &=& \beta_0. \\
\end{array}
\end{displaymath}
The last six terms of the BV structure (\ref{CorrSpaceBVStr}) of the model on the correspondence space then tell us how to define the action of $\Phi$ on the corresponding antifields in such a way that $\Phi$ becomes a BV morphism of $(\mathcal{F}_{E\times_M\hat E},\Omega_{E\times_M\hat E})$, namely
\begin{displaymath}
\begin{array}{rclrcl}
\Phi(\eta^{+0}) &=& \hat{\eta}^+_0, & \Phi(\hat{\eta}^+_0) &=& \eta^{+0}, \\
\Phi(X^+_0) &=& \hat{X}^{+0}, & \Phi(\hat{X}^{+0}) &=& X^+_0, \\
\Phi(\beta^{+0}) &=& \hat{\beta}^+_0, & \Phi(\hat{\beta}^+_0) &=& \beta^{+0}. \\
\end{array}
\end{displaymath}

The swap induced by the Hodge operator (\ref{HodgeSwap}) involves the background fields $A$ and $\hat{A}$, too. Therefore, the BV morphism $\Phi$ should also affect them, namely
\begin{displaymath}
\Phi(A) = \hat{A}, \quad \Phi(\hat{A}) = A.
\end{displaymath}
Together with the action of $\Phi$ on the fiber coordinates $X^0$ and $\hat{X}
_0$, this is equivalent to swapping the connections $\mathcal{A}$ and $\hat{\mathcal{A}}$. 

In effect, $\Phi$ exchanges the two circle bundles $E$ and $\hat{E}$. In other words, it maps the space of fields $\mathcal{F}_{E\times_M\hat E}$ to its dual,
\begin{displaymath}
\Phi: (\mathcal{F}_{E\times_M\hat E},\Omega_{E\times_M\hat E}) \rightarrow(\hat{\mathcal{F}}_{\hat{E}\times_M E},\hat{\Omega}_{\hat{E} \times_M E}).
\end{displaymath}

So far, we considered only the second term of (\ref{clactioncorrsp}). The first term $\int_\Sigma \eta_i \wedge dX^i$ involves only the base manifold $M$ of the circle bundles $E$ and $\hat{E}$ and should therefore be left unaffected by $\Phi$, which is why its action on the corresponding fields $X^i$, $\eta_i$ and $\beta_i$ and their antifields is trivial.

It immediately follows by construction that
\[
\Phi^\ast(\hat{\Omega}_{\hat{E} \times_M E}) = \Omega_{E\times_M\hat E}.
\]

It remains to consider the last term $\int_N X^\ast(h)$, on which the action of $\Phi$ is a bit subtle. Since $\Phi$ affects the background fields $A$ and $\hat{A}$, there is \emph{a priori} no reason why it should leave $h$, the third background field of the model, invariant. Actually, $h$ is the curvature three-form of the basic part $b$ of the B-field introduced in equation (\ref{metricBfieldonE}), $h=db$. As a result, $\Phi$ should follow the T-duality transformation rule (\ref{bhatb}) for this field,
\begin{equation} \label{Phiofh}
\Phi(h) = \Phi(db) = d\hat{b} = h - d(A\wedge \hat{A}) = \hat{h} .
\end{equation}
Note that the application of $\Phi$ follows the prescription for the exchange of the roles of the two circle bundles $E$ and $\hat{E}$ and leaves the invariant part $b_{\mathrm{inv}}$ untouched.

Finally the action of the T-dual model on the correspondence space is the one that fulfills the condition
\begin{displaymath}
\Phi^\ast(\hat{S}_{\hat{E}\times_M E}) = S_{E\times_M\hat E},
\end{displaymath}
namely
\begin{equation}
\begin{split}
\hat{S}_{\hat{E}\times_M E}^{\mathrm{cl}} =& \int_\Sigma \eta_i\wedge dX^i + \left( \hat{\eta}^0 + dX^0 + X^\ast(A) \right) \wedge \left( \eta_0 + d\hat X_0 + X^\ast(\hat A) \right)  \\
& \qquad + \int_N X^\ast(\hat{h}).
\end{split}
\end{equation}

As a side remark, note that in three dimensions, the T-duality morphism exchanged factors in the term $\delta\phi\delta\psi$ of the BV structure, which were symmetric, whereas here it swaps pairs of terms of $\Omega$.
\subsection{The T-dual Effective Model}
In the last step, we look for a dual BV pushforward $\rho'_{\mathrm{UV}\ast}$ that will give us a dual effective action $S^{\mathrm{eff}}_{\hat{E}}$, out of which we will be able to read the dual topological information, namely the dual connection $\hat{\mathcal{A}}$, its curvature $\hat{F}$ and the dual twist $\hat{H}$.

This time, the ultraviolet sector $\hat{\mathcal{F}}_{UV}$ is spanned by
\[
\left(  X^0, \eta_0, \beta_0, \hat{\beta}^0,  X^+_0, \eta^{+0}, \beta^{+0}, \hat{\beta}^+_0 \right)
\]
and is nothing but the image of $\mathcal{F}_{UV}$ under the BV morphism $\Phi$. We choose a similar Lagrangian subspace as before, namely by setting $\hat{\beta}^0=0$, $\beta_0=0$ and $X^0=0$, which keeps the fibers of $\hat{E}$ but not of $E$ as expected. The T-dual effective action we obtain is similar to the original effective action, but the fields associated to the fibers and the background fields have been exchanged,
\begin{equation}
S^{\mathrm{eff}}_{\hat E} =  \int_\Sigma \eta_i\wedge dX^i + \hat{\eta}^0\wedge X^\ast(\hat{\mathcal{A}}) + \eta^{+i}\wedge d\beta_i  + \int_N X^\ast(\hat{h}-\hat{F}\wedge A +  F\wedge\hat{\mathcal{A}}).
\end{equation}
From the WZ term, we find the dual twist
\begin{equation}
\hat{H} = \hat{h}-\hat{F}\wedge A +  F\wedge\hat{\mathcal{A}}
\end{equation}
on the dual circle bundle. From this relation as well as its T-dual (\ref{Hflux1}), we can infer the first two equalities of (\ref{topcontentBuscher}),
\begin{displaymath}
H_{(2)} = \hat F, \quad \hat{H}_{(2)} = F.
\end{displaymath}
We see that these follow directly from the exchange of the two circle bundles in the topological sigma model on the correspondence space with action $S_{E\times_M \hat{E}}$. The change of topology of a circle bundle under a T-duality transformation is thus entirely encoded  in the topological sector of the string sigma model and does not rely on the full set of Buscher rules.

On the other hand, the third equality
\begin{displaymath}
H_{(3)} = \hat{H}_{(3)},
\end{displaymath}
which states that the basic part of the T-dual twist $\hat{H}$ coincides with the one of the original twist $H$, holds only if
\begin{displaymath}
\hat{h}-\hat{F}\wedge A = h - F \wedge \hat{A},
\end{displaymath}
a condition satisfied by the prescription (\ref{Phiofh}) for the action of the BV morphism $\Phi$ on the background fields, which is reminiscent from the Buscher rules. But this actually ensures that the Wess-Zumino consistency condition for the dual effective model, namely $\left[\hat{H}\right] \in H^3(\hat{E}; \mathbb{Z})$, follows from the one on the initial model $\left[H\right] \in H^3(E; \mathbb{Z})$. To see this, one needs to keep in mind that being the curvatures of their dual connections, the vertical parts of the H-fluxes satisfy $\left[H_{(2)}\right] = \left[ \hat{F} \right] \in H^2(M; \mathbb{Z})$ and $\left[\hat{H}_{(2)}\right] = \left[ F \right]  \in H^2(M; \mathbb{Z})$, and that the connections $\mathcal{A}$ and $\hat{\mathcal{A}}$ are normalized to have an integral period around the fibers.

In summary, the topological content of the Buscher rules is encoded in the relation between the topological sigma models with action $S^{\mathrm{eff}}_E$ and $S^{\mathrm{eff}}_{\hat{E}}$, which represent the topological sectors of the involved string sigma models with background fields. Due to the broken $U(1)$ symmetry, a BV morphism could not be readily defined between them, and we had to consider an augmented model on the correspondence space $E\times_M \hat{E}$ to introduce a T-duality transformation, and the two effective models could be retrieved through BV pushforwards, the usual prescription in the BV formalism for effective field theories.
\section{Principal Torus Bundles} \label{TorusBundles}
Our discussion can be generalized to higher dimensional principal torus bundles,
\begin{displaymath}
\begin{CD}
\mathbb{T}^n @>>> E\\
@. @VV\rho V\\
@. M.
\end{CD}
\end{displaymath}
In particular, the connection $\mathcal{A}$ is now a $\mathfrak{t}^n$-valued one-form on $E$, where $\mathfrak{t}^n \simeq \mathbb{R}^n$ is the Lie algebra of $\mathbb{T}^n$, and the corresponding period one generator of the $\mathbb{T}^n$-action $\partial_{\mathcal{A}}$ is $\mathfrak{t}^{\ast n}$-valued.

Before going into technical details, we should analyze once more the results for a circle bundle. In the case of the Courant algebroid, once the spaces of fields for both Courant sigma models based on $(TE\oplus T^\ast E)/S^1$ and $(T\hat{E}\oplus T^\ast \hat{E})/S^1$ were identified with a model space of fields, we were able to interpret the T-duality BV morphism as an automorphism of this model space of fields that left the AKSZ action invariant. In other words, the duality was readily present as a discrete symmetry of the AKSZ action and BV structure, out of which one could find the relations (\ref{topcontentBuscher}). We then applied an adaptation of this morphism to a two-dimensional model. The T-duality was in this case a morphism of the spaces of fields $\mathcal{F}_{E\times_M \hat{E}} \simeq \hat{\mathcal{F}}_{\hat{E} \times_M E}$, but this time the action was not invariant. Nevertheless it still encoded the change of topology induced by the Buscher rules.

For higher-dimensional torus bundles, the procedure is similar. The main difference is that not all H-fluxes allow a T-duality transformation, those are called T-dualizable, and we will see they are precisely the ones that make the Courant sigma model ``T-symmetric''. These T-duality symmetry transformations will form a group, the usual $O(n,n;\mathbb{Z})$ known from string theory. Due to the multiplicity of T-dual models for $n>1$, the construction of morphisms of two-dimensional topological sigma models will be a bit more complicated.

Since the construction of these T-duality BV morphisms for torus bundles closely follows the case of circle bundles, we will not go into all the details, but rather focus on the subtleties implied by a larger T-duality group.
\subsection{Courant Sigma Models}
Given an H-flux $\left[H\right] \in H^3 (E; \mathbb{Z} )$, we can again construct a quotient Courant algebroid on $E$ if we choose an invariant representative $H$ for this cohomology class. Mathematically, it means that $\mathcal{L}_X (H) =0$ for any Killing vector of the torus action on $E$. Moreover, this H-flux is called T-dualizable \cite{Bouwknegt2} if it satisfies the additional requirement that $\imath_X \imath_Y H =0$ for any two Killing vector fields $X$ and $Y$ of the torus action. In practice, this means that the H-flux can be decomposed as
\begin{displaymath}
H = H_{(3)} + \mathcal{A}^a \wedge H_{(2)}^a,
\end{displaymath}
where addition over repeated superscripts $a$ is implicit, $\mathcal{A}^a$, $a=1,\dots,n$, denotes an individual component of the $\mathfrak{t}^n$-valued connection $\mathcal{A}$, and $H_{(3)}$ and $H_{(2)}^a$ are basic three- and two-forms respectively. In particular, no terms quadratic or cubic in the connection $\mathcal{A}$ enters this formula. It appears that $H_{(2)}$ can be interpreted as a $\mathfrak{t}^{\ast n}$-valued two form on the base manifold $M$, and we may write $\mathcal{A}^a \wedge H_{(2)}^a = \mathcal{A} \wedge H_{(2)}$.

The space of fields of the Courant sigma model based on the Courant algebroid $(TE\oplus T^\ast E)/\mathbb{T}^n$ is sensibly the same as the one in the case $n=1$, see equation (\ref{FinvCSM}). A small difference appears for the superfields $\phi$ and $\psi$, which are now defined as
\[
\phi \in \mathrm{Map}( T\left[1\right]N, \mathfrak{t}^n\left[1\right]) \quad \mathrm{and} \quad \psi \in \mathrm{Map}( T\left[1\right]N, \mathfrak{t}^{\ast n}\left[1\right]),
\]
with components $\phi^a,\psi^a \in \mathrm{Fun}(T\left[ 1\right] N,\mathbb{R}\left[ 1\right] )$.

With this notation for the fields, the AKSZ BV structure and action remain the same as in the case $n=1$, but now (\ref{CSMBVstrE}) and (\ref{CSMactionE}) really mean
\begin{displaymath}
\Omega = \int_{T\left[1\right]N} \mu\, \left( \delta p_i \, \delta X^i - \delta\xi^i \, \delta\Theta_i - \delta \phi^a \,\delta\psi^a \right).
\end{displaymath}
and
\begin{equation} \label{CSMtorusaction1}
\begin{split}
S = \int_{T\left[1\right]N}\mu & \, \left( p_i \, DX^i + \frac{1}{2}\xi^i \, D\Theta_i + \frac{1}{2}\Theta_i \, D\xi^i + \frac{1}{2}\phi^a \,D\psi^a + \frac{1}{2}\psi^a \, D\phi^a \right. \\
& \left. -p_i\xi^i + \psi^a\,\frac{1}{2}F^a_{ij}\xi^i\xi^j + \frac{1}{6}H_{(3)ijk}\xi^i\xi^j\xi^k + \phi^a\,\frac{1}{2}H^a_{(2)ij}\xi^i\xi^j \right).
\end{split}
\end{equation}

From now on, we will identify $\mathfrak{t}^n \simeq \mathfrak{t}^{\ast n} \simeq \mathbb{R}^n$ with $\mathbb{R}^n$. We can therefore combine the $\mathbb{R}^{n}\left[1\right]$-valued superfields $\phi$ and $\psi$ into a $\mathbb{R}^{2n}\left[1\right]$-valued superfield
\begin{displaymath}
\Xi = \left( \begin{array}{c} \phi \\ \psi \end{array} \right) \in \mathrm{Map}( T\left[1\right]N, \mathbb{R}^{2n}\left[1\right]).
\end{displaymath}
This allows us to re-write the last term in the AKSZ BV structure $\Omega$ as
\begin{displaymath}
\delta \phi^a \, \delta \psi^a
= \frac{1}{2} \left( \delta\phi \ \delta\psi \right) \left( \begin{array}{cc} 0 & \mathbf{1} \\ \mathbf{1} & 0 \end{array} \right) \left( \begin{array}{c} \delta\phi \\ \delta\psi \end{array} \right)
= \frac{1}{2} \, \delta\Xi^T \, K \, \delta\Xi = \frac{1}{2}\langle \delta\Xi,\delta\Xi\rangle_K,
\end{displaymath}
where $K = \left( \begin{array}{cc} 0 & \mathbf{1} \\ \mathbf{1} & 0 \end{array} \right)$ is a $2n\times 2n$ symmetric matrix used to define the scalar product $\langle v,w\rangle_K = v^T  \, K \, w$. We can thus write the BV structure
\begin{equation}
\Omega = \int_{T\left[1\right]N} \mu\, \left( \delta p_i \, \delta X^i - \delta\xi^i \, \delta\Theta_i - \frac{1}{2}\langle \delta\Xi,\delta\Xi\rangle_K \right).
\end{equation}

We see that $\Omega$ is invariant under linear transformations of the $\Xi$ superfield, 
\begin{displaymath}
\Xi \mapsto \mathcal{O} \Xi = \hat{\Xi},
\end{displaymath}
such that $\mathcal{O}^T \, K \, \mathcal{O}=\mathbf{1}$. Since $K$ can be diagonalized to $\left( \begin{array}{cc} \mathbf{1} & 0 \\ 0 & -\mathbf{1} \end{array} \right)$, this transformations form the group $O(n,n;\mathbb{R})$. Such a linear transformation can be lifted to a BV automorphism
\begin{displaymath}
\Phi_{\mathcal{O}} : \mathcal{F} \rightarrow \mathcal{F}.
\end{displaymath}

To see how it affects the action $S$, it is best to combine the $\mathbb{R}^n$-valued two-forms $F$ and $H_{(2)}$ into an $\mathbb{R}^{2n}$-valued two-form
\begin{displaymath}
\mathbf{F} = \left( \begin{array}{c} F \\ H_{(2)} \end{array} \right),
\end{displaymath}
so that one can write the BV action
\begin{equation}
\begin{split} \label{CSMtorusaction2}
S = \int_{T\left[1\right]N}\mu & \, \left( p_i \, DX^i + \frac{1}{2}\xi^i \, D\Theta_i + \frac{1}{2}\Theta_i \, D\xi^i + \frac{1}{2}\langle \Xi, D\Xi \rangle_K \right. \\
& \left. -p_i\xi^i + \langle \Xi , \frac{1}{2}\mathbf{F}_{ij}\xi^i\xi^j \rangle_K + \frac{1}{6}H_{(3)ijk}\xi^i\xi^j\xi^k \right).
\end{split}
\end{equation}
By comparison with the circle bundle case, we see that we can construct a dual action
\begin{equation}
\begin{split}
\hat{S} = \int_{T\left[1\right]N}\mu & \, \left( p_i \, DX^i + \frac{1}{2}\xi^i \, D\Theta_i + \frac{1}{2}\Theta_i \, D\xi^i + \frac{1}{2}\langle \hat{\Xi}, D\hat{\Xi} \rangle_K \right. \\
& \left. -p_i\xi^i + \langle \hat{\Xi} , \frac{1}{2}\hat{\mathbf{F}}_{ij}\xi^i\xi^j \rangle_K + \frac{1}{6}H_{(3)ijk}\xi^i\xi^j\xi^k \right).
\end{split}
\end{equation}
that satisfies the duality requirement $\Phi_{\mathcal{O}}^\ast (\hat{S}) = S$ provided $\Phi_{\mathcal{O}}(\mathbf{F}) = \hat{\mathbf{F}}$, and that we have actually $\hat{S}=S$ if $\Phi_{\mathcal{O}}(\mathbf{F}) = \mathcal{O} \mathbf{F}$, so that the Courant algebroids encoded in the Courant sigma models actions $S$ and $\hat{S}$ are actually isomorphic. 

It is essential to note that the $2n$ components of $\mathbf{F}$ define integral cohomology classes,
\[
\left[ F^a \right], \ \left[ H^a_{(2)} \right] \in H^2(M;\mathbb{Z}), \ a=1,\dots,n,
\]
because the curvature of a circle bundle has integral periods and $H$ needs to satisfy the Wess-Zumino consistency condition. The $2n$ components of $\hat{\mathbf{F}}$ are subject to the same constraint. This is ensured if $\mathcal{O}$ is taken in the subgroup $O(n,n;\mathbb{Z})$ of $O(n,n;\mathbb{R})$. This is actually the T-duality group for the torus $T^n$, which leads us to interpret $\Phi_{\mathcal{O}}$ as a transformation of the fibers of the torus bundle $E$ into a dual bundle $\Phi_{\mathcal{O}}(E)$.

This formulation of toroidal T-duality in the Courant sigma model has the additional advantage to cast some new light on the condition of the twist $H$ to be T-dualizable. Had $H$ not satisfied this condition, so would the Courant sigma model action for the torus bundle (\ref{CSMtorusaction1}) have contained terms quadratic or cubic in $\phi$, and we would not have been able to re-write it in the symmetric form (\ref{CSMtorusaction2}).
\subsection{Two-dimensional Sigma Models}
A similar duality between two-dimensional models as in section \ref{TdualityPSM} is a bit more complicated to work out for higher-dimensional tori. When $n=1$, we have $O(1,1;\mathbb{Z}) = \mathbb{Z}_2$, which makes the T-dual of a certain circle bundle with twist unique. We used this unicity to combine the connections associated to both bundles in a generalization of the Poisson sigma model action, on which the T-duality group $\mathbb{Z}_2$ acted by exchange of the two bundles. For $n>1$, the group is larger, but $\left\lbrace \mathbf{1},K \right\rbrace$ is a canonical $\mathbb{Z}_2$ subgroup of $O(n,n;\mathbb{Z})$ that we may use to pick a specific dual principal circle bundle.

As in the case $n=1$, given a principal $\mathbb{T}^n$-bundle $E$ with T-dualizable twist $H$, we start with a classical action
\begin{equation} \label{initialactionontorusbundle}
S^{\mathrm{cl}}_E = \int_\Sigma \eta_i\wedge dX^i + \theta \wedge X^\ast(\mathcal{A}) + \int_N X^\ast(H),
\end{equation}
where $\mathcal{A}$ is a connection on $E$, therefore a $\mathfrak{t}^n$-valued one-form on $E$. This implies in turn that $\theta$ (which replaces $\eta_0$ from the $n=1$ case) is a $\mathfrak{t}^{\ast n}$-valued one-form on $\Sigma$.

Like in three dimensions, we cast the curvature $F$ of the connection $\mathcal{A}$ and the component $H_{(2)}$ into an $\mathbb{R}^{2n}$-valued two-form $\mathbf{F}$. We can interpret $H_{(2)}$ as the curvature of the connection $\hat{\mathcal{A}}$ of a dual torus bundle $\hat{E}$. In that case, $\mathbf{F}$ corresponds to the curvature of the connection
\begin{displaymath}
\mathbf{A} = \left( \begin{array}{c} \mathcal{A} \\ \hat{\mathcal{A}} \end{array} \right)
\end{displaymath}
of the $\mathbb{T}^{2n}$-bundle $E \times_M \hat{E}$. To gauge the whole $\mathbb{T}^n \times \mathbb{T}^n$ symmetry, we need to introduce a connection $\hat{\theta}$ (a generalization of $\hat{\eta}^0$), locally a $\mathfrak{t}^n$-valued one-form on $\Sigma$. We may regroup it with $\theta$ into an $\mathbb{R}^{2n}$-valued connection
\[
\Theta = \left( \begin{array}{c} \hat{\theta} \\ \theta \end{array} \right).
\]
With the matrix $J = \left( \begin{array}{cc} 0 & -\mathbf{1} \\ \mathbf{1} & 0 \end{array} \right)$, we can write an action on the correspondence space that generalizes (\ref{clactioncorrsp}),
\begin{equation}
S^{\mathrm{cl}}_{E \times_M \hat{E}} = \int_\Sigma \left( \eta_i \wedge dX^i + \frac{1}{2}\left( \Theta + X^\ast(\mathbf{A})\right)^T \wedge J \left( \Theta + X^\ast(\mathbf{A})\right) \right) + \int_N X^\ast (h).
\end{equation}
The basic twist $h$ is a generalization of (\ref{Hflux1}) for the $n=1$ situation,
\begin{displaymath}
h = H + d(\mathcal{A}^a \wedge \hat{A}^a)
\end{displaymath}
where $\hat{A}$ is the gauge potential associated to the connection $\hat{\mathcal{A}}$.

We write only the classical part of the action, the full BV action can easily be inferred from its version (\ref{CorrSpaceBVAct}) for circle bundles.

To retrieve the original action associated to the topological sector of a string on the torus bundle $E$, we treat as UV degrees of freedom the fiber coordinates of the torus bundle $\hat{E}$, the first half of $\Theta$ and of course all the ghosts. The functional integration still yields a constant, and the effective action on the IR sector is our initial action (\ref{initialactionontorusbundle}), as expected, with $d(\mathcal{A} \wedge \hat{A})$ completing $h$ to the whole twist $H$.

The $O(n,n;\mathbb{Z})$ T-duality group acts on $\mathbf{A}$ and $\Theta$ by matrix multiplication. For $n>1$, we evidently obtain more than one T-dual model. In this case, each
\begin{displaymath}
\mathcal{O}\mathbf{F} = \mathbf{F}' = \left( \begin{array}{c} F' \\ H_{(2)}' \end{array} \right)
\end{displaymath}
will determine a pair of torus bundles $E'$ and $\hat{E}'$, and the BV morphism $\Phi_{\mathcal{O}}$ associated to the T-duality transformation $\mathcal{O} \in O(n,n;\mathbb{Z})$ maps the space of fields of the topological model on the correspondence space $E\times_M \hat{E}$ to the one of the model on $E' \times_M \hat{E}'$,
\begin{displaymath}
\Phi_{\mathcal{O}} : \mathcal{F}_{E\times_M \hat{E}} \rightarrow \mathcal{F}_{E'\times_M \hat{E}'}. 
\end{displaymath}
It is also required to map $h$ to $h' = \Phi_{\mathcal{O}}(h)$ in such a way that the basic part of the twist of the effective models based on the torus bundles $E$ and $E'$ is invariant. The idea is to use the fact that $h=db$ and to adapt the transformation (\ref{bhatb}), but instead of just exchanging $A$ with $\hat{A}$, we can form an $\mathbb{R}^{2n}$-valued gauge potential $(A, \hat{A})^T$, act on it with $\mathcal{O}$,
\begin{displaymath}
\left( \begin{array}{c} A' \\ \hat{A}' \end{array} \right) = \mathcal{O} \left( \begin{array}{c} A \\ \hat{A} \end{array} \right),
\end{displaymath}
and find
\begin{displaymath}
h' = h + \frac{1}{2}d( A' \wedge \hat{A}' - A \wedge \hat{A}).
\end{displaymath}

The process to find the effective model based on $E'$ is similar as the one for $E$. One starts with the action $S_{E' \times_M \hat{E}'}$ and chooses the UV sector to be made of the fiber coordinates of $\hat{E}'$, the first half of $\Theta' = \mathcal{O}\Theta$ and the ghosts.

Note that if we choose for $\mathcal{O}$ the particular element $K = \left( \begin{array}{cc} 0 & \mathbf{1} \\ \mathbf{1} & 0 \end{array} \right)$, we obtain the same swap of the bundles $E$ and $\hat{E}$ that we had in the case of circle bundles, namely $E' = \hat{E}$ and $\hat{E}' = E$. This was to be expected, as $K$ is the only non-trivial element of $O(1,1;\mathbb{Z}) = \mathbb{Z}_2$.

To summarize, we saw that in the case of principal torus bundles, the T-duality group acts on topological field theories by BV morphisms. For three-dimensional TFTs, we recovered isomorphisms of Courant algebroids when the H-flux was T-dualizable. In two dimensions, through BV pushforwards from the various T-dual models, we were able to find the topological sectors of the T-dual string sigma models with background fields associated to these principal torus bundles and their relations corresponding to the topological content of the Buscher rules.


\begin{thebibliography}{99}
\bibitem{AKSZconstruction} M. Alexandrov, M. Kontsevich, A. Schwarz, and O. Zaboronsky, ``The geometry of the Master equation and topological quantum field theory'', Int. J. Modern Phys. A, 12(7):1405–1429, 1997
\bibitem{BVtheorem} I. Batalin and G. Vilkovisky, ``Gauge algebra and quantization'', Phys. Lett., 102B:27, 1981
\bibitem{Bouwknegt1} P. Bouwknegt, J. Evslin, and V. Mathai, ``T-duality: topology change from H- flux'', Comm. Math. Phys., 249(2):383–415, 2004, hep-th/0306062
\bibitem{Bouwknegt2}  P. Bouwknegt, K. Hannabuss, and V. Mathai, ``T-duality for principal torus bundles'',  J. High Energy Phys., 3:018, 10 pp. (electronic), 2004, hep-th/0312284
\bibitem{Bouwknegt3} P. Bouwknegt, K. Hannabuss, V. Mathai, `` T-duality for principal torus bundles and dimensionally reduced Gysin sequences'', Adv. Theor. Math. Phys. 9:749-773, 2005,  	arXiv:hep-th/0412268
\bibitem{Buscherrules1} T. H. Buscher, ``A symmetry of the string background field equations'', Phys. Lett. B, 194(1):59–62, 1987
\bibitem{Buscherrules2} T. H. Buscher, ``Path-integral derivation of quantum duality in nonlinear sigma- models'', Phys. Lett. B, 201(4):466–472, 1988
\bibitem{AKSZPSM} A. S. Cattaneo, G. Felder, `` On the AKSZ formulation of the Poisson sigma model'', Lett.Math.Phys. 56 (2001) 163-179,  	arXiv:math/0102108
\bibitem{BVBFV} A. S. Cattaneo, P. Mnev, N. Reshetikhin, ``Classical BV theories on manifolds with boundary'',  	arXiv:1201.0290
\bibitem{TdualityCourant} G. R. Cavalcanti, M. Gualtieri, ``Generalized complex geometry and T-duality'', A Celebration of the Mathematical Legacy of Raoul Bott (CRM Proceedings \& Lecture Notes), American Mathematical Society, 2010, pp. 341-366. ISBN: 0821847775, 	arXiv:1106.1747 [math.DG]
\bibitem{HoriVafa} K. Hori, C. Vafa, ``Mirror Symmetry'', hep-th/0002222
\bibitem{Kao} P. Kao, ``T-duality and Poisson-Lie T-duality in generalized geometry'', (2008)
\bibitem{WZPSM} C. Klim\v c\' ik, T. Strobl, ``WZW-Poisson manifolds'', J. Geom. Phys. 43 (2002), no. 4, 341–344
\bibitem{CourantAlgebroid} Z.-J. Liu, A. Weinstein, and P. Xu, ``Manin triples for Lie Bialgebroids'', Journ. of Diff.geom. 45 pp.647–574 (1997)
\bibitem{LosevEffective} A. Losev, talk at GAP, Perugia, 2005
\bibitem{effectiveBV} P. Mnev, ``Discrete BF theory'', arXiv:0809.1160
%\bibitem{MnevAKSZObs} P. Mnev, `` A construction of observables for AKSZ sigma models'',  	arXiv:1212.5751
\bibitem{AKSZCSM} D. Roytenberg, ``AKSZ-BV Formalism and Courant Algebroid-induced Topological Field Theories'', Lett. Math. Phys. 79:143-159, 2007
\bibitem{PSMfirst} P. Schaller, T. Strobl, ``Poisson structure induced (topological) field theories'', Modern Phys. Lett. A 9 (1994), no. 33, 3129–3136
\bibitem{SchwarzGeomQME} A. Schwarz, ``Geometry of Batalin-Vilkovisky quantization'', Commun.Math.Phys. 155 (1993) 249-260
\bibitem{PavolClass} P. \v Severa, ``Letters to A. Weinstein'', unpublished
\end{thebibliography}
\end{document}